\documentclass{article}
\pdfoutput=1
\usepackage{jcappub}
\usepackage{graphicx}
\usepackage{url}
\usepackage[center]{subfigure}
\usepackage{caption}
\usepackage{xspace}
\usepackage{footnote}
\usepackage{amssymb}
\usepackage{textcomp}

\allowdisplaybreaks	

\newcommand{\red}{\color{red}}
\definecolor{purple}{rgb}{0.5,0,0.5}
\newcommand{\purple}{\color{purple}}
\definecolor{darkgreen}{rgb}{0.1,0.8,0.1}
\newcommand{\green}{\color{darkgreen}}

\newcommand{\fn}[1]{\footnotesize{#1}}
\newcommand{\FAIL}{\red\large\texttimes}
\newcommand{\CHECK}{\large\green\checkmark}

\begin{document}


\vspace*{-15mm}
\begin{flushright}
MPP-2017-78
\end{flushright}
\vspace*{0.7cm}

\title{``Non-cold'' dark matter at small scales: a general approach}

\author[a,b]{R.~Murgia}
\emailAdd{riccardo.murgia@sissa.it}

\author[c]{A.~Merle}
\emailAdd{amerle@mpp.mpg.de}

\author[a,d,b]{M.~Viel}
\emailAdd{viel@sissa.it}

\author[c]{M.~Totzauer}
\emailAdd{totzauer@mpp.mpg.de}

\author[e]{A.~Schneider}
\emailAdd{aurel.schneider@phys.ethz.ch}

\affiliation[a]{SISSA, via Bonomea 265, 34136 Trieste, Italy}
\affiliation[b]{INFN, Sezione di Trieste, via Valerio 2, I-34127 Trieste, Italy}
\affiliation[c]{Max Planck Institute for Physics, F\"ohringer Ring 6, 80805 Munich, Germany}
\affiliation[d]{INAF/OATS, Osservatorio Astronomico di Trieste, via Tiepolo 11, I-34143 Trieste, Italy}
\affiliation[e]{Institute for Astronomy, Department of Physics, ETH Zurich, Wolfgang-Pauli-Strasse 27, 8093, Zurich, Switzerland}


\abstract{Structure formation at small cosmological scales provides an important frontier for dark matter (DM) research. Scenarios with small DM particle masses, large momenta or hidden interactions tend to suppress the gravitational clustering at small scales. The details of this suppression depend on the DM particle nature, allowing for a direct link between DM models and astrophysical observations. However, most of the astrophysical constraints obtained so far refer to a very specific shape of the power suppression, corresponding to thermal warm dark matter (WDM), i.e., candidates with a Fermi-Dirac or Bose-Einstein momentum distribution. In this work we introduce a new analytical fitting formula for the power spectrum, which is simple yet flexible enough to reproduce the clustering signal of large classes of non-thermal DM models, which are not at all adequately described by the oversimplified notion of WDM. We show that the formula is able to fully cover the parameter space of sterile neutrinos (whether resonantly produced or from particle decay), mixed cold and warm models, fuzzy dark matter, as well as other models suggested by effective theory of structure formation (ETHOS). Based on this fitting formula, we perform a large suite of $N$-body simulations and we extract important nonlinear statistics, such as the matter power spectrum and the halo mass function. Finally, we present first preliminary astrophysical constraints, based on linear theory, from both the number of Milky Way satellites and the Lyman-$\alpha$ forest. This paper is a first step towards a general and comprehensive modeling of small-scale departures from the standard cold DM model.}

\maketitle

\section{\label{sec:intro}Introduction}

Cosmic microwave background and large-scale structure data have provided the consistent picture that the present universe is mainly composed by a cosmological constant, denoted by $\Lambda$, and by cold dark matter (CDM). Whereas this \emph{standard cosmological model}, or $\Lambda$CDM model, provides very convincing predictions for the large-scale structures, it still exhibits some limits at very small sub-galactic scales (for a review see~\cite{weinberg}), where tensions exist between predictions and observations. In fact, assuming the $\Lambda$CDM model, cosmological $N$-body simulations predict too many dwarf galaxies (\emph{missing satellite} problem~\cite{Klypin:1999uc,Moore:1999nt}) and too much dark matter (DM) in the innermost regions of galaxies (\emph{cusp-core} problem~\cite{2010AdAst2010E...5D}) with respect to observations. The situation both under the theoretical/numerical and observational point of view is not entirely clear, and there are continuous efforts to exploit astrophysical consequences of DM both in (spiral) galaxies~\cite{donato09,salucci13}, dwarf galaxies of our Milky Way (MW), or, even more recently, tidal streams~\cite{belokurov}. Moreover, the dynamical properties of massive MW satellites are not reproduced in simulations (\emph{too-big-to-fail} problem~\cite{2011MNRAS415L40B,2012MNRAS4221203B}). These small-scale problems could also be alleviated or solved by invoking baryon physics, still not perfectly understood and/or fully implemented in cosmological simulations. For instance, it is known that photo-evaporation from UV sources during the reionisation period pushes gas out from small halos, preventing star formation and reducing the number of observed substructures~\cite{Okamoto:2008sn}. Moreover, supernova feedback may be able to make the inner parts of halo density profiles significantly shallower~\cite{Governato:2012fa}. However, these baryonic feedback effects are currently implemented as semi-analytical models into hydrodynamical simulations, making it difficult to obtain results with full predictive power.

Another possible solution to the small-scale crisis of the $\Lambda$CDM model is to modify the nature of DM, by going beyond the standard CDM paradigm. In fact, despite huge efforts both in particle physics and cosmology, the nature and composition of the ``dark sector'' are still unknown. From an astrophysical perspective, DM candidates can be classified according to either a velocity dispersion or a pressure term which counteracts the effects of gravity at small scales. Below this characteristic scale of a given DM candidate, density fluctuations are erased and gravitational clustering is thus suppressed. The velocity dispersion of CDM particles is by definition so small to be totally negligible for cosmological structure formation.

Many ``non-cold'' DM (nCDM) candidates, well motivated by particle physics theories, such as sterile neutrinos~\cite{Adhikari:2016bei,Konig:2016dzg,Schneider:2016uqi,Merle:2017jfn,yeche17,devega} or axion-like particles~\cite{Hu:2000ke,Marsh:2013ywa,Hui:2016ltb,irsic17fuzzy,armengaud17}, have been proposed in order to give a better description of the structure formation and distribution at small scales, with respect to the $\Lambda$CDM model. In addition, there are other non-standard hypotheses potentially able to induce a small-scale suppression in the matter power spectrum: a mixed (cold and warm) DM fluid~\cite{Schneider:2016ayw,Viel2005,diamanti,Gariazzo:2017pzb}, DM particles coupled to dark energy~\cite{Wang:2016lxa,Murgia:2016ccp,Murgia:2016dug} or to some relativistic fluid~\cite{Boehm:2014vja,Bringmann:2016ilk}, or Self-Interacting Dark Matter (SIDM)~\cite{Cyr-Racine:2015ihg,Vogelsberger:2015gpr}. Different scenarios lead to different shapes in the suppression of the power spectrum. However, most of the constraints from structure formation data which have been published so far, refer to a very specific shape of the small-scale power suppression, corresponding to the case of {\emph{thermal}} warm dark matter (WDM), i.e., candidates with a Fermi-Dirac momentum distribution~\cite{Viel:2013apy,Baur:2015jsy,Lapi:2015zea,Irsic:2017ixq}. Nonetheless, most of viable nCDM candidates do not have thermally distributed momenta (i.e., they feature truly \emph{non-thermal} distribution functions), which may lead to non-trivial suppressions in their power spectra. In particular, given that non-thermal distributions may feature more than one characteristic momentum scale, these settings may in fact provide a whole new approach to resolving the small scale issues of CDM.

The suppression of gravitational clustering in nCDM models is usually parametrised through the so-called \emph{transfer function} $T(k)$, i.e., the square root of the ratio of the matter power spectrum predicted by the given model with respect to that in the presence of CDM only, for fixed cosmological parameters. In this work we introduce a new, more general analytical fitting formula for the transfer function. We show that it  accurately describes the behaviour of the most popular DM scenarios, being able to reproduce a large variety of shapes with only three free parameters. Finally, we present the first constraints from structure formation on its free parameters.

This paper is organised as follows: in Sec.~\ref{sec:param} we introduce and motivate our general approach; in Sec.~\ref{sec:models} we show that the new formula is an useful tool for fitting the behaviour of the most viable DM models provided by particle physics; in Sec.~\ref{sec:results} we discuss the results of the cosmological simulations that we have performed in order to investigate different parametrisations of the new transfer function; in Sec.~\ref{sec:obs} we present the first preliminary astrophysical constraints on its free parameters; finally, in Sec.~\ref{sec:reality-check}, we apply these constraints to the theoretical particle physics models previously discussed, in order to compare the predictions in terms of structure formation by our fits with the actual transfer functions and check
the accuracy of our method.

\section{\label{sec:param}Method and parametrisation}

\begin{figure}[t]
  \centering
  \includegraphics[width=12.5cm]{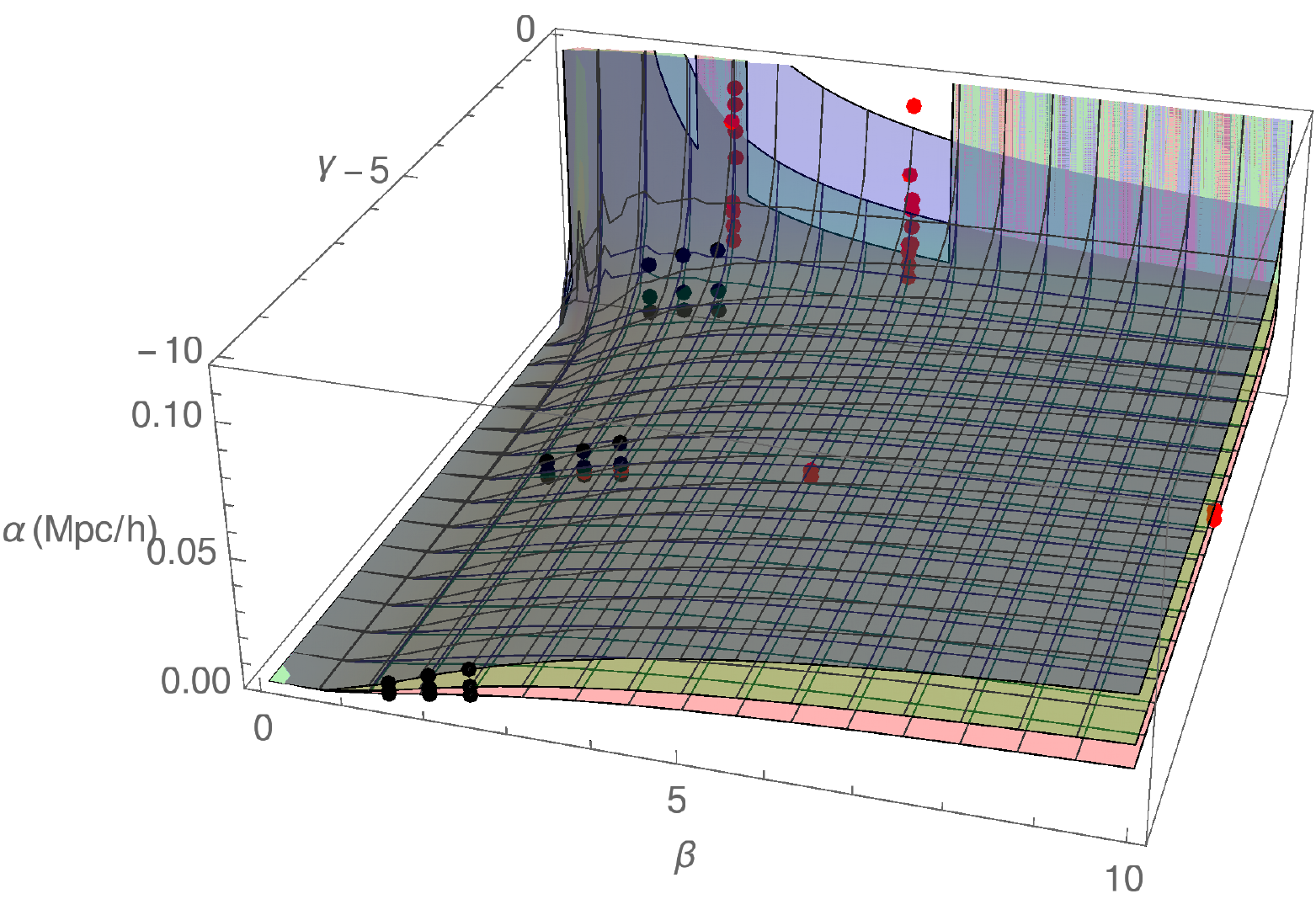}
  \caption{\label{fig:grid}The blue, green, and red surfaces represent the regions of the $\{\alpha,\beta,\gamma\}$ space corresponding to thermal WDM masses of 2, 3, and 4~keV, respectively. The black dots constitute the $3\times 3\times 3$ non-regular grid that we have considered, while the red dots correspond to the additional 28 points that we have taken into account for our analyses (see Tab.~\ref{tab:params}).}
\end{figure}

\begin{figure}[b]
   \hspace{-0.5cm}
  \begin{tabular}{lr}
  \includegraphics[width=7.8cm]{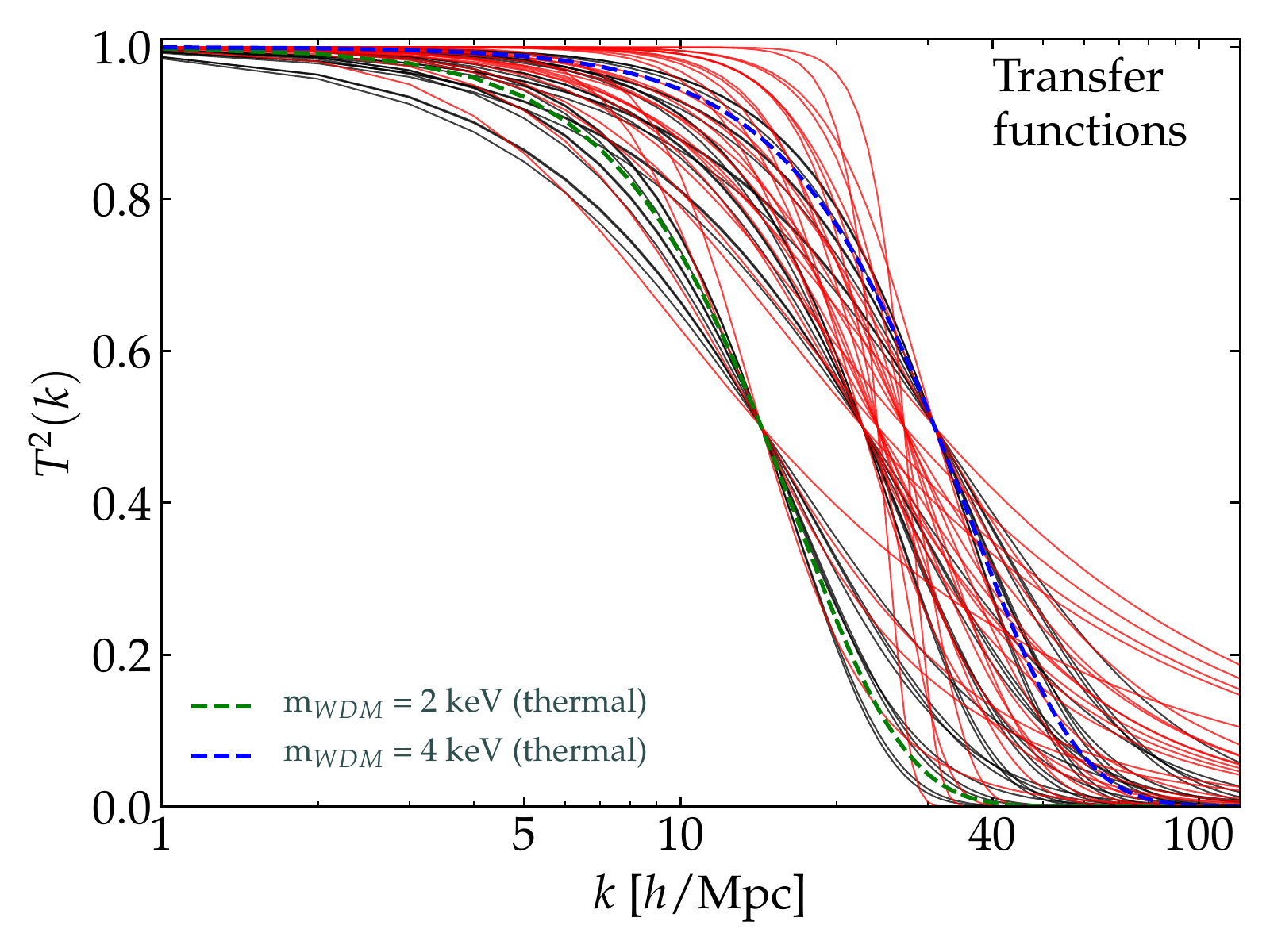} & \includegraphics[width=7.8cm]{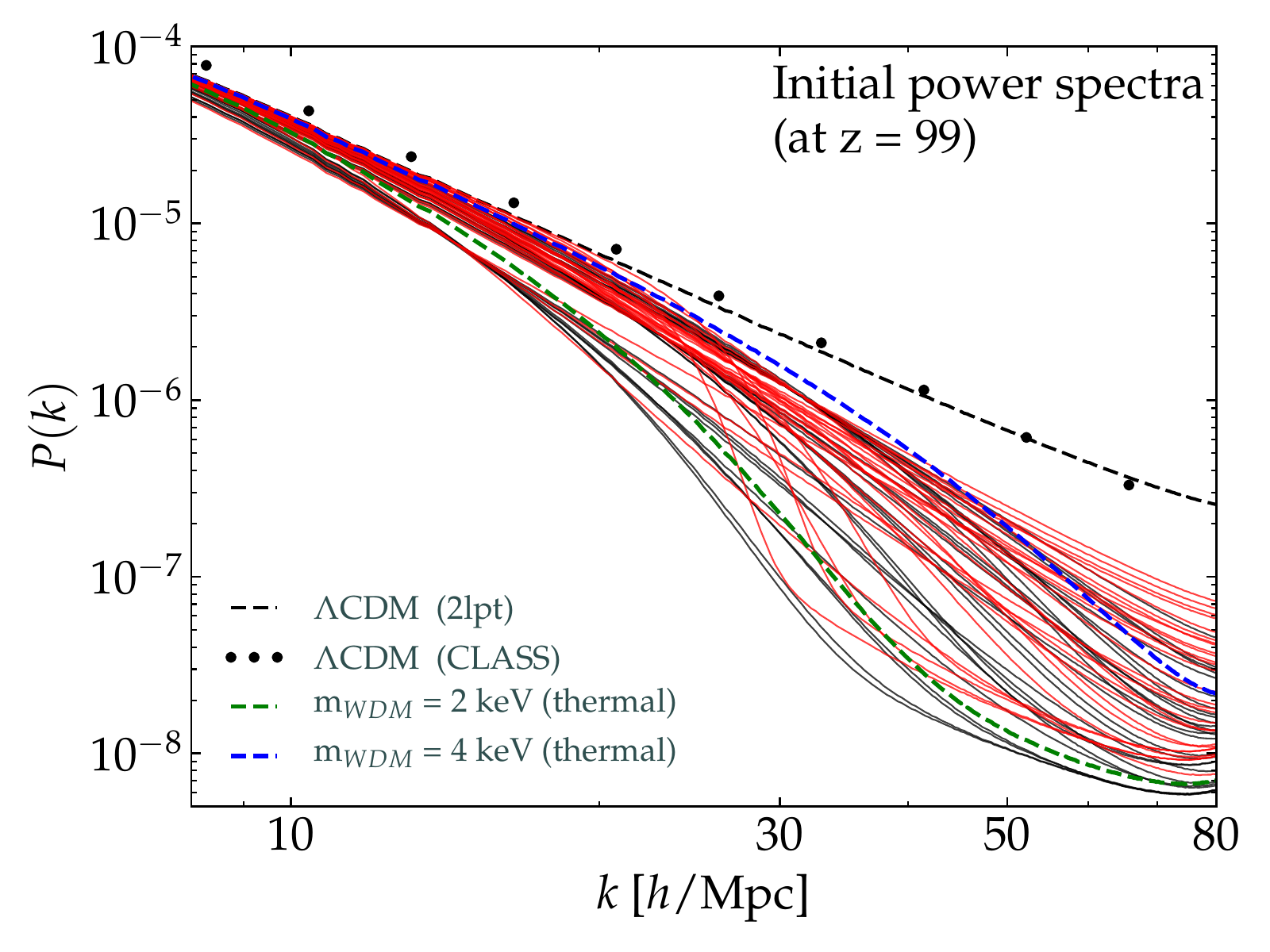}
  \end{tabular}
  \caption{\label{fig:Tk}
  {\emph{Left panel}}:~the black solid lines are the transfer functions computed by Eq.~\eqref{eq:Tgen} and associated to the 27~combinations of $\{\alpha,\beta,\gamma\}$ listed in Tab.~\ref{tab:params}, whereas the red solid lines correspond to the additional 28~triplets that we have considered for our analyses. The green and blue dashed lines are the transfer functions computed in Eq.~\eqref{eq:Viel} for thermal WDM masses of 2 and 4~keV, respectively. {\emph{Right panel}}:~the black solid lines are the total matter power spectra, computed at redshift $z=99$, associated to the first 27 nCDM models listed in Tab.~\ref{tab:params}, while the red solid lines refer to the last 28 cases, listed in Tab.~\ref{tab:params} as well. The black dashed line corresponds to the $\Lambda$CDM case. The dotted line correspond to the $\Lambda$CDM case, computed in linear theory. The green and blue dashed lines refer to thermal WDM models with masses of 2 and 4~keV, respectively.}
\end{figure}

The suppression of gravitational clustering in nCDM scenarios can be described through the transfer function $T(k)$, given by:
\begin{equation}\label{eq:Tkdef}
 T^2(k) \equiv \frac{P_{\rm{nCDM}}(k)}{P_{\rm{CDM}}(k)},
\end{equation}
where $P_{\rm CDM}$ and $P_{\rm nCDM}$ are the power spectra of the CDM and the nCDM models, respectively.

For the special case of thermal WDM, the transfer function can be well approximated by the fitting function~\cite{Bode2001}:
\begin{equation}\label{eq:Viel}
 T(k) = [ 1 + (\alpha k)^{2\nu} ]^{-5/\nu},
\end{equation}
where $\alpha$ is the only free parameter and $\nu = 1.12$. Therefore, constraints on the mass of the WDM candidate translate into bounds on $\alpha$, by the following formula~\cite{Viel2005}:
\begin{align}
 \begin{split}
 \alpha &= 0.24 \left( \frac{m_x/T_x}{1~\rm{keV}/T_\nu} \right)^{-0.83} \left( \frac{\omega_x}{0.25\cdot (0.7)^2} \right)^{-0.16}{\rm{Mpc}}\\
 &= 0.049 \left( \frac{m_x}{1~\rm{keV}} \right)^{-1.11} \left( \frac{\Omega_x}{0.25} \right)^{0.11} \left(\frac{h}{0.7}\right)^{1.22}h^{-1}\rm{Mpc}~,
 \end{split}
 \label{eq:alphaold}
\end{align}
where the subscripts $x$ and $\nu$ refer to WDM and active-neutrino properties, respectively, and the second equation holds only in the case of thermal relics.

Let us now generalize Eq.~\eqref{eq:Viel} and write down the following fitting formula\footnote{Note that equivalent fitting functions have already been used in, e.g., Refs.~\cite{Barkana:2001gr,viel12,Destri:2013hha}. However, they have only been applied to special cases, and its general applicability had not been recognised to our knowledge.}
\begin{equation}\label{eq:Tgen}
 T(k) = [ 1 + (\alpha k)^{\beta} ]^{\gamma},
\end{equation}
which is a function of three free parameters: $\alpha$, $\beta$, and $\gamma$.
In this paper we show that the simple function given by Eq.~\eqref{eq:Tgen} is generic enough to describe the majority of nCDM models from the literature.

As a next step, let us define the characteristic half-mode scale $k_{1/2}$, obtained by setting $T^2\equiv 1/2$. Using Eq.~\eqref{eq:Tgen}, we therefore have:
\begin{equation}\label{eq:k12}
 k_{1/2} = \frac{1}{\alpha} \left[ \left( \frac{1}{\sqrt{2}} \right)^{1/\gamma} -1\right]^{1/\beta}.
\end{equation}
Whereas through Eqs.~\eqref{eq:Viel} and~\eqref{eq:alphaold} we had a one-to-one correspondence between $m_x$ and $\alpha$, constraints on the DM mass are now, by Eqs.~\eqref{eq:Tgen} and \eqref{eq:k12}, mapped to 3-dimensional surfaces in the $\{\alpha,\beta,\gamma\}$ space. In other words, given a value of $k_{1/2}$, which corresponds to a certain (thermal) WDM mass, we can easily compute the corresponding surface in the 3-dimensional parameter space from Eq.~\eqref{eq:k12} -- but this information alone is not yet sufficient to decide about the validity of the point under consideration. In Fig.~\ref{fig:grid} we plot the three surfaces associated to the $k_{1/2}$ values listed below: 

\begin{align}
\begin{split}
k'_{1/2}=14.323 ~h/{\rm{Mpc}} {~~~{\rm (}\emph{if thermal}{\rm :}~\longleftrightarrow~ m'_x = 2~\rm{keV}{\rm ),}} \\
k''_{1/2}=22.463 ~h/{\rm{Mpc}} {~~~{\rm (}\emph{if thermal}{\rm :}~\longleftrightarrow~ m''_x = 3~\rm{keV}{\rm ),}} \\
k'''_{1/2}=30.914 ~h/{\rm{Mpc}} {~~~{\rm (}\emph{if thermal}{\rm :}~\longleftrightarrow~ m'''_x = 4~\rm{keV}{\rm ).}}
\end{split}        
\label{eq:kvals}        
\end{align}
We also build up a $3\times 3\times 3$ non-regular grid that brackets the volume of the parameter space between the blue upper surface (corresponding to $k'_{1/2}$) and the red lower surface (corresponding to $k'''_{1/2}$) in Fig.~\ref{fig:grid}. This is done by taking all the possible combinations of the two triplets $\beta=\{1.5,2,2.5\}$ and $\gamma=\{-1,-5,-10\}$ and computing the corresponding values of $\alpha$, by plugging the values of $k_{1/2}$ listed in Eq.~\eqref{eq:kvals} into Eq.~\eqref{eq:k12}.

\begin{table}[t]
 \begin{center}
  \begin{tabular}{|c|c|c|c|c|}
  \hline
				   &  \small{$\beta$}	& \small{$\gamma$}	& \small{$\alpha$~\scriptsize{(Mpc/$h$)}}& \small{$k_{1/2}$~\scriptsize{($h$/Mpc)}}	   \\ \hline
\scriptsize{nCDM1}	   &  \footnotesize{1.5} 	&\footnotesize{$-10.0$}     & \footnotesize{0.0075}  & \footnotesize{14.323}    \\ 
\scriptsize{nCDM2}	   &  \footnotesize{1.5} 	&\footnotesize{$-10.0$}     & \footnotesize{0.0048}  & \footnotesize{22.463}    \\ 
\scriptsize{nCDM3}	   &  \footnotesize{1.5} 	&\footnotesize{$-10.0$}     & \footnotesize{0.0035}  & \footnotesize{30.914}    \\ 
\scriptsize{nCDM4}	   &  \footnotesize{1.5} 	&\footnotesize{$-5.0$}     & \footnotesize{0.012 }  & \footnotesize{14.323}   \\ 
\scriptsize{nCDM5}	   &  \footnotesize{1.5} 	&\footnotesize{$-5.0$}     & \footnotesize{0.0077}  & \footnotesize{22.463}    \\ 
\scriptsize{nCDM6}	   &  \footnotesize{1.5} 	&\footnotesize{$-5.0$}     & \footnotesize{0.0056}  & \footnotesize{30.914}    \\ 
\scriptsize{nCDM7}	   &  \footnotesize{1.5} 	&\footnotesize{$-1.0$}     & \footnotesize{0.039 }  & \footnotesize{14.323}   \\ 
\scriptsize{nCDM8}	   &  \footnotesize{1.5} 	&\footnotesize{$-1.0$}     & \footnotesize{0.025 }  & \footnotesize{22.463}   \\ 
\scriptsize{nCDM9}	   &  \footnotesize{1.5} 	&\footnotesize{$-1.0$}     & \footnotesize{0.018 }  & \footnotesize{30.914}   \\
\scriptsize{nCDM10}  &  \footnotesize{2.0} 	&\footnotesize{$-10.0$}     & \footnotesize{0.013 }  & \footnotesize{14.323}   \\ 
\scriptsize{nCDM11}  &  \footnotesize{2.0} 	&\footnotesize{$-10.0$}     & \footnotesize{0.0084}  & \footnotesize{22.463}    \\ 
\scriptsize{nCDM12}  &  \footnotesize{2.0} 	&\footnotesize{$-10.0$}     & \footnotesize{0.0061}  & \footnotesize{30.914}    \\ 
\scriptsize{nCDM13}  &  \footnotesize{2.0} 	&\footnotesize{$-5.0$}     & \footnotesize{0.019 }  & \footnotesize{14.323}   \\ 
\scriptsize{nCDM14}  &  \footnotesize{2.0} 	&\footnotesize{$-5.0$}     & \footnotesize{0.012 }  & \footnotesize{22.463}   \\ 
\scriptsize{nCDM16}  &  \footnotesize{2.0} 	&\footnotesize{$-1.0$}     & \footnotesize{0.045 }  & \footnotesize{14.323}   \\ 
\scriptsize{nCDM15}  &  \footnotesize{2.0} 	&\footnotesize{$-5.0$}     & \footnotesize{0.0087}  & \footnotesize{30.914}    \\ 
\scriptsize{nCDM17}  &  \footnotesize{2.0} 	&\footnotesize{$-1.0$}     & \footnotesize{0.029 }  & \footnotesize{22.463}   \\ 
\scriptsize{nCDM18}  &  \footnotesize{2.0} 	&\footnotesize{$-1.0$}     & \footnotesize{0.021 }  & \footnotesize{30.914}   \\ 
\scriptsize{nCDM19}  &  \footnotesize{2.5} 	&\footnotesize{$-10.0$}     & \footnotesize{0.018 }  & \footnotesize{14.323} \\
\scriptsize{nCDM20}  &  \footnotesize{2.5} 	&\footnotesize{$-10.0$}     &\footnotesize{0.012 }   &\footnotesize{22.463}    \\ 
\scriptsize{nCDM21}  &  \footnotesize{2.5} 	&\footnotesize{$-10.0$}     &\footnotesize{0.0085}   &\footnotesize{30.914}     \\ 
\scriptsize{nCDM22}  &  \footnotesize{2.5} 	&\footnotesize{$-5.0$}      &\footnotesize{0.024 }   &\footnotesize{14.323}    \\ 
\scriptsize{nCDM23}  &  \footnotesize{2.5} 	&\footnotesize{$-5.0$}      &\footnotesize{0.016 }   &\footnotesize{22.463}    \\ 
\scriptsize{nCDM24}  &  \footnotesize{2.5} 	&\footnotesize{$-5.0$}      &\footnotesize{0.011 }   &\footnotesize{30.914}    \\ 
\scriptsize{nCDM25}  &  \footnotesize{2.5} 	&\footnotesize{$-1.0$}      &\footnotesize{0.049 }   &\footnotesize{14.323}    \\ 
\scriptsize{nCDM26}  &  \footnotesize{2.5} 	&\footnotesize{$-1.0$}      &\footnotesize{0.031 }   &\footnotesize{22.463}    \\ 
\scriptsize{nCDM27}  &  \footnotesize{2.5} 	&\footnotesize{$-1.0$}      &\footnotesize{0.023 }   &\footnotesize{30.914}    \\
                  &                     &                        &                      &                       \\ \hline
\end{tabular}
  \begin{tabular}{|c|c|c|c|c|}
  \hline
				   &  \small{$\beta$}	& \small{$\gamma$}	& \small{$\alpha$~\scriptsize{(Mpc/$h$)}}& \small{$k_{1/2}$~\scriptsize{($h$/Mpc)}}	   \\ \hline
\red\scriptsize{nCDM28}  & \red\footnotesize{2}      &\red\footnotesize{$-5$}        &\red\footnotesize{0.011}    &\red\footnotesize{24.0}\\ 
\red\scriptsize{nCDM29}  & \red\footnotesize{2}      &\red\footnotesize{$-5$}        &\red\footnotesize{0.0099}   &\red\footnotesize{27.0}\\
\red\scriptsize{nCDM30}  & \red\footnotesize{2.5}     &\red\footnotesize{$-5$}        &\red\footnotesize{0.015}    &\red\footnotesize{24.0}\\ 
\red\scriptsize{nCDM31}  & \red\footnotesize{2.5}     &\red\footnotesize{$-5$}        &\red\footnotesize{0.013}    &\red\footnotesize{27.0}\\ 
\red\scriptsize{nCDM32}  & \red\footnotesize{5}      &\red\footnotesize{$-5$}        &\red\footnotesize{0.025}    &\red\footnotesize{24.0}\\ 
\red\scriptsize{nCDM33}  & \red\footnotesize{5}      &\red\footnotesize{$-5$}        &\red\footnotesize{0.022}    &\red\footnotesize{27.0}\\ 
\red\scriptsize{nCDM34}  & \red\footnotesize{10}      &\red\footnotesize{$-5$}        &\red\footnotesize{0.032}    &\red\footnotesize{24.0}\\ 
\red\scriptsize{nCDM35}  & \red\footnotesize{10}      &\red\footnotesize{$-5$}        &\red\footnotesize{0.029}    &\red\footnotesize{27.0}\\
\red\scriptsize{nCDM36}  & \red\footnotesize{2.5}     &\red\footnotesize{$-0.3$}       &\red\footnotesize{0.095}    &\red\footnotesize{14.323}\\
\red\scriptsize{nCDM37}  & \red\footnotesize{2.5}     &\red\footnotesize{$-0.15$}      &\red\footnotesize{0.17}     &\red\footnotesize{14.323}\\
\red\scriptsize{nCDM38}  & \red\footnotesize{2.5}     &\red\footnotesize{$-0.3$}       &\red\footnotesize{0.061}    &\red\footnotesize{22.463}\\          
\red\scriptsize{nCDM39}  & \red\footnotesize{2.5}     &\red\footnotesize{$-0.15$}    &\red\footnotesize{0.11}   &\red\footnotesize{22.463}\\
\red\scriptsize{nCDM40}  & \red\footnotesize{2.5}     &\red\footnotesize{$-0.3$}     &\red\footnotesize{0.044}  &\red\footnotesize{30.914}\\
\red\scriptsize{nCDM41}  & \red\footnotesize{2.5}     &\red\footnotesize{$-0.15$}    &\red\footnotesize{0.078}  &\red\footnotesize{30.914}\\
\red\scriptsize{nCDM42}  & \red\footnotesize{2.5}     &\red\footnotesize{$-0.3$}     &\red\footnotesize{0.057}  &\red\footnotesize{24.0}\\
\red\scriptsize{nCDM43}  & \red\footnotesize{2.5}     &\red\footnotesize{$-0.15$}    &\red\footnotesize{0.10}   &\red\footnotesize{24.0}\\
\red\scriptsize{nCDM44}  & \red\footnotesize{2.5}     &\red\footnotesize{$-0.3$}     &\red\footnotesize{0.051}  &\red\footnotesize{27.0}\\
\red\scriptsize{nCDM45}  & \red\footnotesize{2.5}     &\red\footnotesize{$-0.15$}    &\red\footnotesize{0.090}  &\red\footnotesize{27.0}\\
\red\scriptsize{nCDM46}  & \red\footnotesize{5}	        &\red\footnotesize{$-0.3$}     &\red\footnotesize{0.082}  &\red\footnotesize{14.323}\\
\red\scriptsize{nCDM47}  & \red\footnotesize{5}	        &\red\footnotesize{$-0.15$}    &\red\footnotesize{0.11}   &\red\footnotesize{14.323}\\
\red\scriptsize{nCDM48}  & \red\footnotesize{5}	        &\red\footnotesize{$-0.3$}     &\red\footnotesize{0.052}  &\red\footnotesize{22.463}\\
\red\scriptsize{nCDM49}  & \red\footnotesize{5}	        &\red\footnotesize{$-0.15$}    &\red\footnotesize{0.069}  &\red\footnotesize{22.463}\\
\red\scriptsize{nCDM50}  & \red\footnotesize{5}	        &\red\footnotesize{$-0.3$}     &\red\footnotesize{0.038}  &\red\footnotesize{30.914}\\
\red\scriptsize{nCDM51}  & \red\footnotesize{5}	        &\red\footnotesize{$-0.15$}    &\red\footnotesize{0.050}  &\red\footnotesize{30.914}\\
\red\scriptsize{nCDM52}  & \red\footnotesize{5}	        &\red\footnotesize{$-0.3$}     &\red\footnotesize{0.049}  &\red\footnotesize{24.0}\\
\red\scriptsize{nCDM53}  & \red\footnotesize{5}	        &\red\footnotesize{$-0.15$}    &\red\footnotesize{0.065}  &\red\footnotesize{24.0}\\
\red\scriptsize{nCDM54}  & \red\footnotesize{5}	        &\red\footnotesize{$-0.3$}     &\red\footnotesize{0.043}  &\red\footnotesize{27.0}\\
\red\scriptsize{nCDM55}  & \red\footnotesize{5}	        &\red\footnotesize{$-0.15$}    &\red\footnotesize{0.058}  &\red\footnotesize{27.0}\\\hline
\end{tabular}
\caption{\label{tab:params}Each $\{\alpha,\beta,\gamma\}$-combination corresponds to a different nCDM model. The first 27~models are associated to the combinations of $\{\alpha,\beta,\gamma\}$ that constitute the $3\times 3\times 3$ non-regular grid in the parameter space, while the last 28~models refer to the additional triplets that we have considered. In the last column we list the corresponding values of $k_{1/2}$.}
\end{center}
\end{table}

Thus, we obtain a table with 27~combinations of $\{\alpha,\beta,\gamma\}$ that sample the volume of the parameter space associated to thermal WDM masses between 2 and 4~keV, each of them corresponding to a different nCDM model. The models are listed in Tab.~\ref{tab:params}, with labels from 1~to~27, and are represented in Fig.~\ref{fig:grid} by the black points. In order to investigate even shallower and steeper transfer functions, we have furthermore considered 28 additional points in the parameter space, corresponding to the $\{\alpha,\beta,\gamma\}$-combinations listed in Tab.~\ref{tab:params}, with labels from 28~to~55, represented by the red points in Fig.~\ref{fig:grid}. Note that, although the points marked in Fig.~\ref{fig:grid} may appear to be somewhat sparcely distributed at first sight, they in fact cover a large fraction of the relevant parameter space. The reason for this lies in a quasi-degeneracy between the two parameters $\alpha$ and $\gamma$,
which we discuss in detail in Appendix~\ref{ap:degeneracy}.

For each of the models listed in Tab.~\ref{tab:params}, we have computed the corresponding transfer function by using Eq.~\eqref{eq:Tgen}. We plot them in the left panel of Fig.~\ref{fig:Tk}, where the green and blue dashed lines represent the ``old'' transfer functions, i.e.\ computed through Eq.~\eqref{eq:Viel}, for $m'_x=2 ~\rm{keV}$ and $m'''_x=4 ~\rm{keV}$, respectively.~Fig.~\ref{fig:Tk} gives an overview over the large variety of transfer functions investigated in this paper.

Let us now qualitatively describe the role of the different parameters in the generalised fit for the transfer function, Eq.~\eqref{eq:Tgen}. The value of $\alpha$ gives the general scale of suppression, i.e., it is the most important parameter for setting the position of $k_{1/2}$. The parameters $\beta$ and $\gamma$ are responsible for the slope of the transfer function before and after the half-mode scale $k_{1/2}$, respectively. The parameter $\beta$ has to be positive in order to have meaningful transfer functions, since $\beta < 0$ gives a $T(k)$ that differs from 1 at large scales. The larger is $\beta$, the flatter is the transfer function before $k_{1/2}$. Analogously, the larger the absolute value of $\gamma$, the sharper is the cut-off.

\section{\label{sec:models}Connection to particle physics models}

The purpose of this section is to see to which extent the suggested 3-parameter fitting formula is able to match the transfer functions from different nCDM models. We will focus on sterile neutrinos from resonant production, sterile neutrinos from particle decay production, mixed (cold plus warm) DM models, ultra-light axions, and another class of models suggested by the effective theory of structure formation (ETHOS).

\subsection{\label{sec:models_Resonant}Sterile neutrinos by resonant production}

\looseness=-1 Given that keV sterile neutrinos generically mix with the active-neutrino sector, it is a natural idea to use this mixing to produce sterile neutrino DM in the early universe.
\looseness=-1 While it is nowadays known that the production by non-resonant transitions (``Dodelson-Widrow mechanism'')~\cite{Langacker:1989sv,Dodelson:1993je,Merle:2015vzu}\footnote{Note that, contrary to previous statements in the literature~\cite{Dodelson:1993je,Colombi:1995ze}, non-resonantly produced sterile neutrinos also feature a non-thermal distribution~\cite{Merle:2015vzu}, rather than a suppressed thermal spectrum. However, while this slightly changes the published numerical values of the bounds on this setting, the basic conclusion of non-resonant production being excluded remains valid (in fact, it is even made stronger)~\cite{Merle:2015vzu}.} is incompatible with structure formation~\cite{Viel:2013apy,Adhikari:2016bei}, a suitable lepton number asymmetry in the early universe (whose origin is not necessarily clear, though) can resonantly enhance the active-sterile transitions (``Shi-Fuller mechanism'') and yield spectra that are more likely to be in agreement with data~\cite{Enqvist:1990ek,Shi:1998km,Abazajian:2001nj,Canetti:2012kh,Venumadhav:2015pla,Ghiglieri:2015jua} (note, however, that also this mechanism is restricted to a small successful region in the parameter space~\cite{Cherry:2017dwu,Adhikari:2016bei}).

\begin{figure}[t]
  \hspace{-0.5cm}
  \begin{tabular}{lr}
  \includegraphics[width=7.9cm]{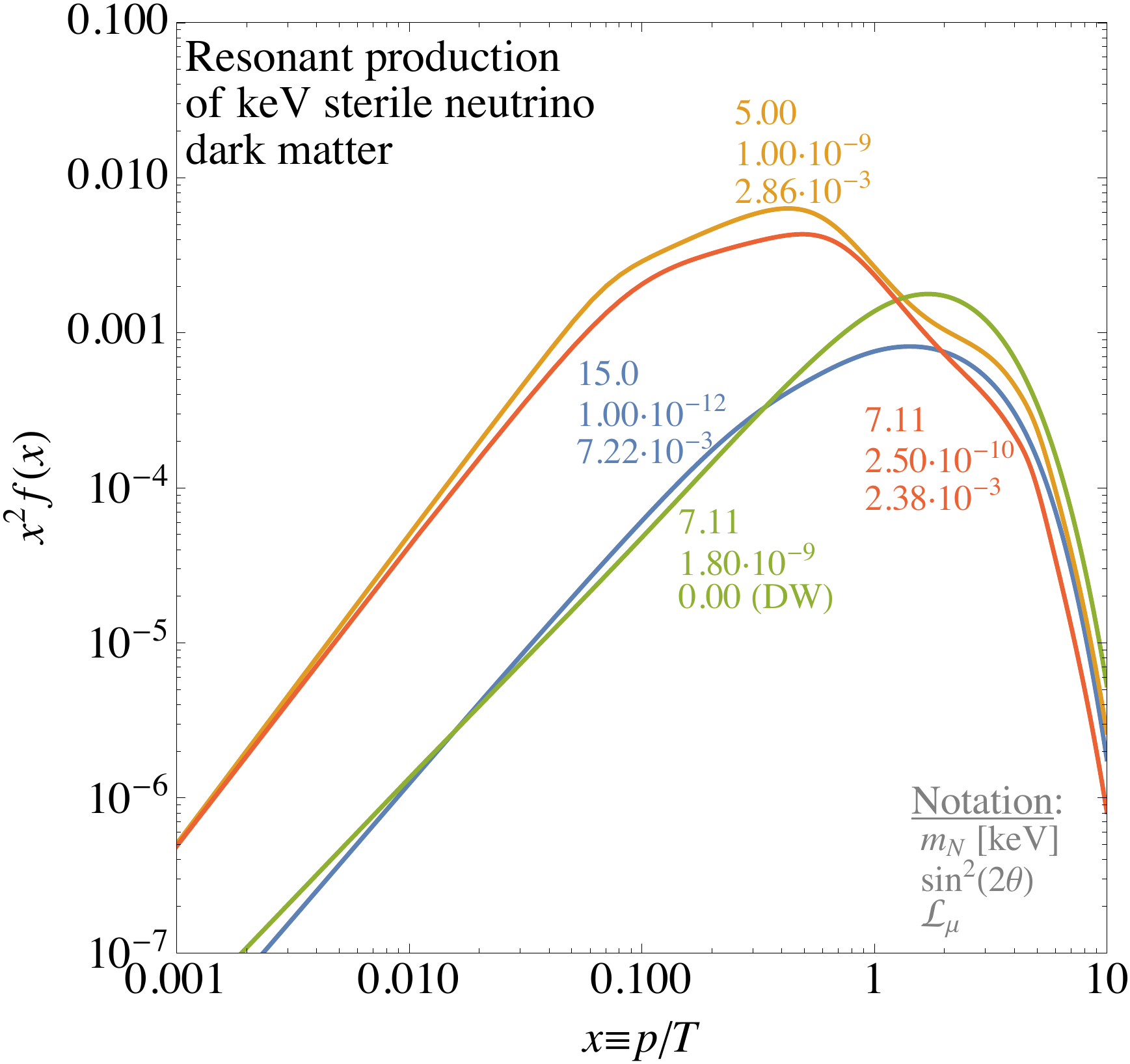} & \includegraphics[width=7.5cm]{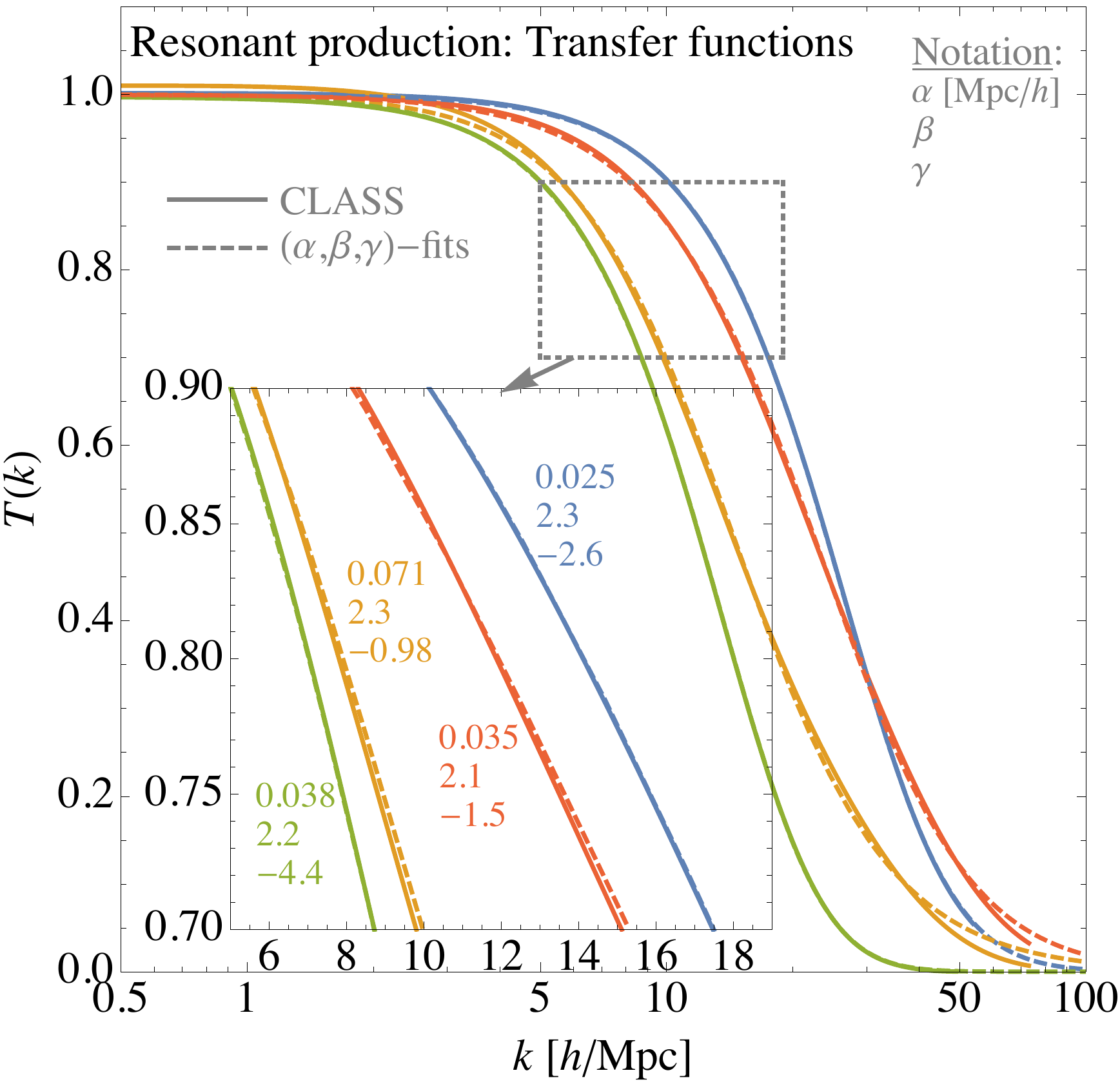}
  \end{tabular}
  \caption{\label{fig:Res-fits} Example distributions functions (\emph{left}) and corresponding transfer functions (at $z=0$; \emph{right}) for resonant (Shi-Fuller) production, including one example for non-resonant (Dodelson-Widrow) production, green curve, for comparison. On the right panel, one can see that the transfer functions are fitted very well by the parametrisation from Eq.~\eqref{eq:Tgen}.}
\end{figure}

As can be seen from Refs.~\cite{Abazajian:2001nj,Venumadhav:2015pla,Ghiglieri:2015jua}, the distribution functions resulting from resonant production can be highly non-thermal: typically, they feature one or more narrow peaks on top of a continuous spectrum, see Fig.~\ref{fig:Res-fits}. This figure shows different momentum distribution functions (\emph{left}) and the corresponding transfer functions (\emph{right}), for a few example values of the sterile neutrino mass $m_N$, the active-sterile mixing angle $\sin^2 (2\theta)$, and the lepton asymmetry $\mathcal{L}_\mu$.\footnote{Due to the current technical limitations of the software developed in conjunction with Ref.~\cite{Venumadhav:2015pla}, the lepton asymmetry can only be placed in the muon sector, if the package {\tt sterile-dm} is used. However, as shown in Ref.~\cite{Ghiglieri:2015jua}, the results would not be altered dramatically if the lepton asymmetry was present in another sector.} Note that the green curve actually features $\mathcal{L}_\mu \equiv 0$, i.e., a case of non-resonant production. Compared to the red curve, one can see that in this case a larger mixing angle is required to meet the correct abundance, and also the spectrum is different from the resonant cases. The plots in the right panel illustrate the corresponding transfer functions (solid lines), along with the fits obtained from Eq.~\eqref{eq:Tgen} using a least squares approach. As can be seen already by eye (and is confirmed by a goodness-of-fit test), our general transfer function, Eq.~\eqref{eq:Tgen}, provides excellent fits to these cases, with parameter values well within the parameter space probed in this work, cf.\ Fig.~\ref{fig:grid}. This remains true also for distributions with more than one scale, which can e.g.\ be seen for the orange curve in the plots. Indeed, the trick of fitting the transfer function gets us rid of many redundancies.

\looseness=-1 Given that the distributions we tested (more than shown here) are very representative for resonant production of sterile neutrino DM, 
we conclude that our fit function describes this class of models very well.

\subsection{\label{sec:models_Decay}Sterile neutrinos from particle decays}

\begin{figure}[t]
  \hspace{-0.5cm}
  \begin{tabular}{lr}
  \includegraphics[width=7.9cm]{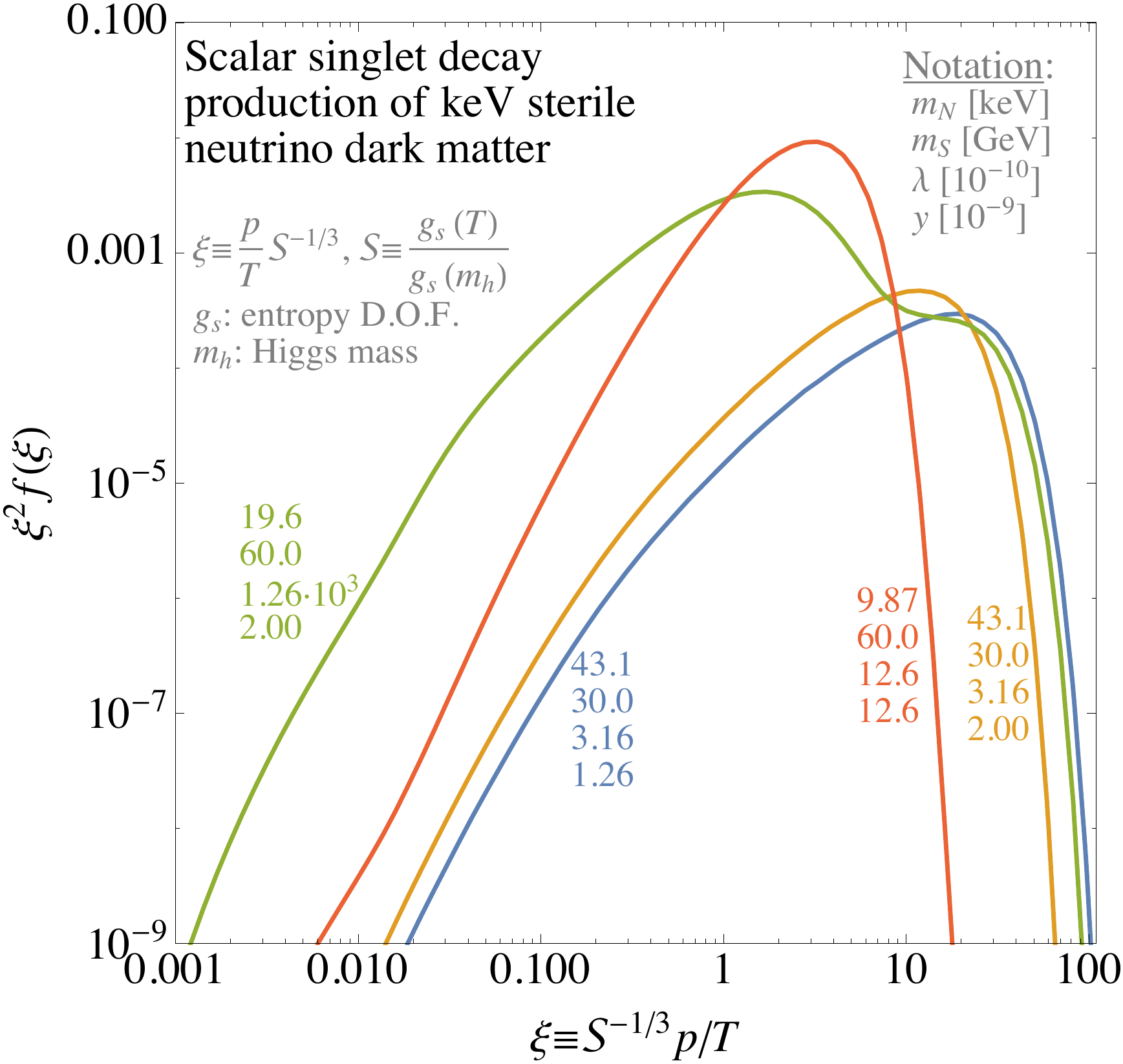} & \includegraphics[width=7.6cm]{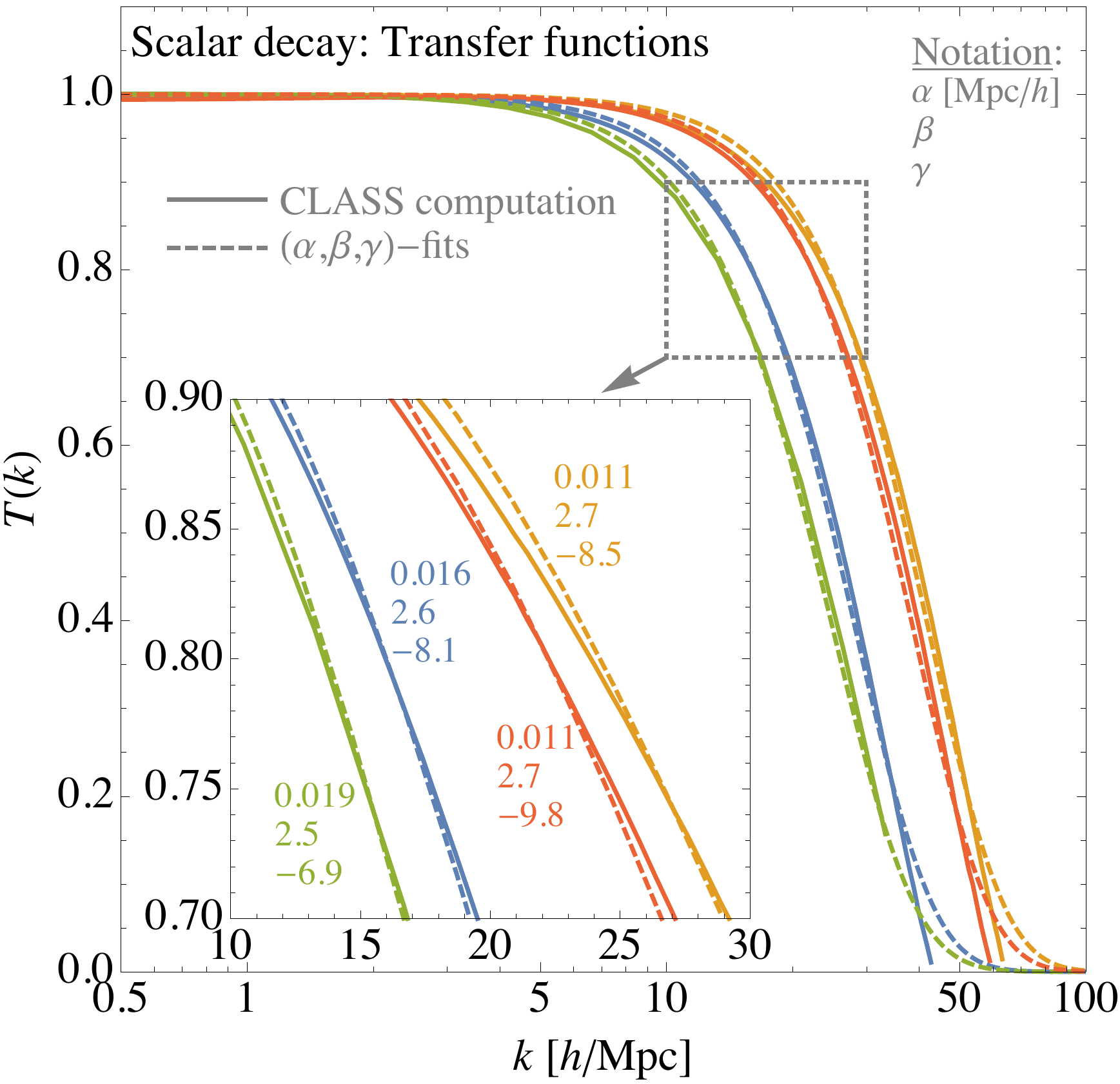}
  \end{tabular}
  \caption{\label{fig:SD-fits}Example distributions functions (\emph{left}) and corresponding transfer functions (at $z=0$; \emph{right}) for scalar decay production. On the right panel, one can see that the transfer functions are fitted very well by the parametrisation from Eq.~\eqref{eq:Tgen}.}
\end{figure}

Another potential production mechanism for sterile neutrino DM relies on the decay of a hypothetical parent particle in the early universe, whose properties (in terms of the momentum distribution) translate into those of the resulting keV sterile neutrino. A very simple case discussed frequently in the literature is that of a singlet scalar particle which may via its interactions with the Standard Model Higgs boson either thermalise (and thus be equilibrated and ultimately freeze-out) or not (and thus freeze-in), see Refs.~\cite{Kusenko:2006rh,Petraki:2007gq,Merle:2013wta,Merle:2015oja,Shakya:2015xnx,Konig:2016dzg} for very detailed treatments. Other possibilities for parent particles, like pions~\cite{Lello:2014yha} or electrically charged scalars~\cite{Frigerio:2014ifa}, do not exhibit any behaviour that would be qualitatively different.

For decay production, the resulting distribution functions are highly non-thermal, with spectra not only having a shape very different from a thermal one but also featuring, in general, two distinct characteristic scales~\cite{Merle:2015oja,Konig:2016dzg} (or even three if a subdominant subsequent Dodelson-Widrow modification is taken into account~\cite{Merle:2015vzu}). Four example distributions are depicted in the left panel of Fig.~\ref{fig:SD-fits}, for different values of the sterile neutrino and decaying scalar masses, along with the two parameters $\lambda$ (Higgs portal) and $y$ (Yukawa coupling), which shape the distributions -- see Refs.~\cite{Merle:2015oja,Konig:2016dzg} for details. The plots in the right panel illustrate the corresponding transfer functions (solid lines), along with the fits obtained from Eq.~\eqref{eq:Tgen} using a least squares approach. As for the case of resonantly produced sterile neutrinos, the general parametrisation of Eq.~\eqref{eq:Tgen}, provides an excellent fit with parameter values well within the range studied in this paper (cf.\ Fig.~\ref{fig:grid}). This remains true also for distributions with more than one scale, which can be seen for the green curve in the plots. Indeed, the trick of fitting the transfer function gets us rid of many redundancies.

Given that the distributions we tested (more than shown here) are very representative for decay production of DM, independently of the details of the particle physics setting under consideration, we conclude that our fit function describes this class of models very well.

\subsection{\label{sec:models_Mixed}Mixed models}

\looseness=-1 In principle, the DM sector may consist of a complicated mixture of different DM particles~\cite{Boyarsky:2008xj}. Here we study a toy model that assumes the presence of both a cold and a warm (thermal) component.
\looseness=-1 This simple model, dubbed mixed DM, leads to a large variety of shapes in the transfer function, therefore providing an ideal test for the parametrisation used in this paper (i.e.\ the fitting formula of Eq.~\eqref{eq:Tgen}).

\begin{figure}[t]
  \hspace{-0.5cm}
  \begin{tabular}{lr}
  \includegraphics[width=7.9cm]{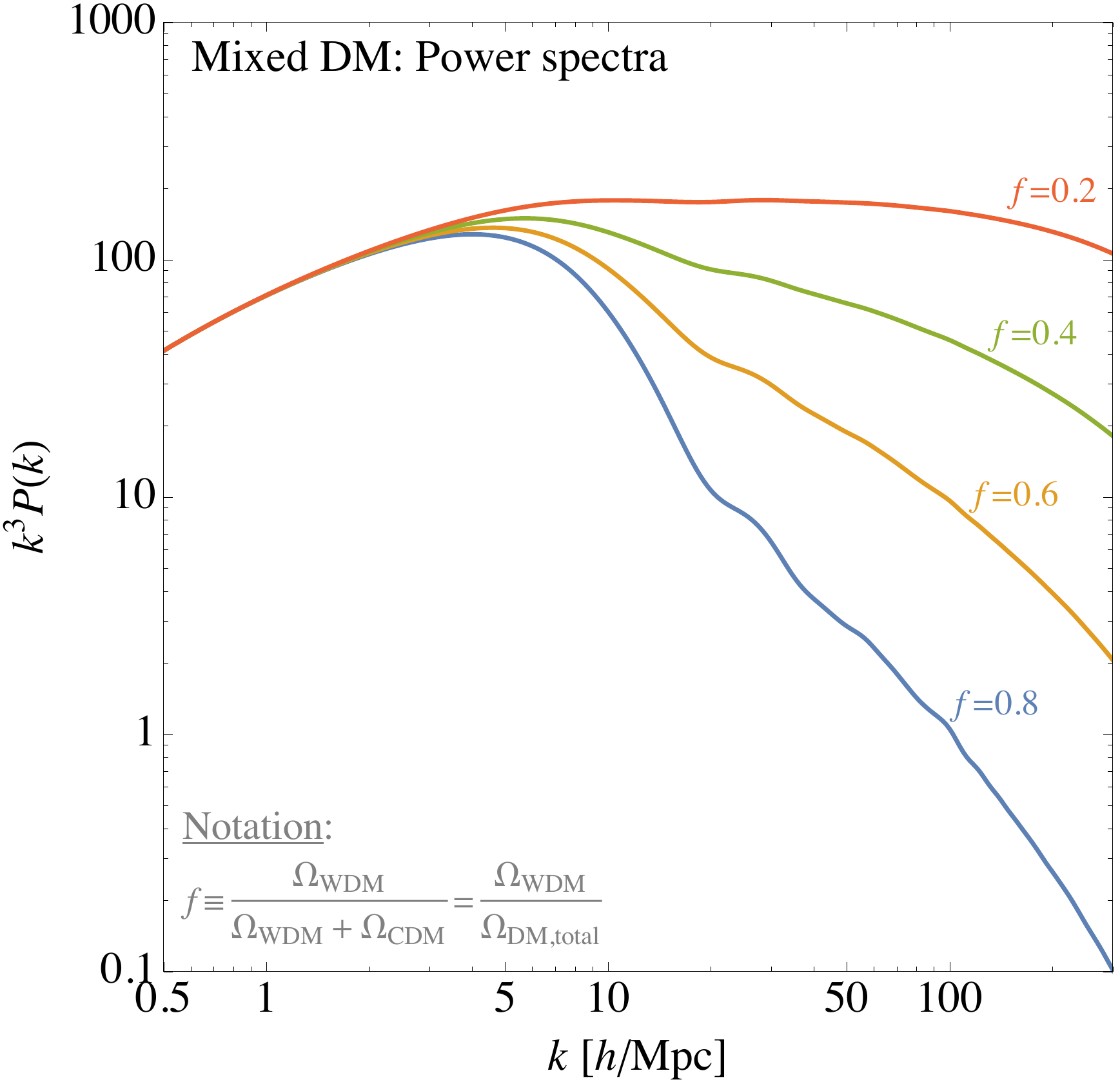} & \includegraphics[width=7.75cm]{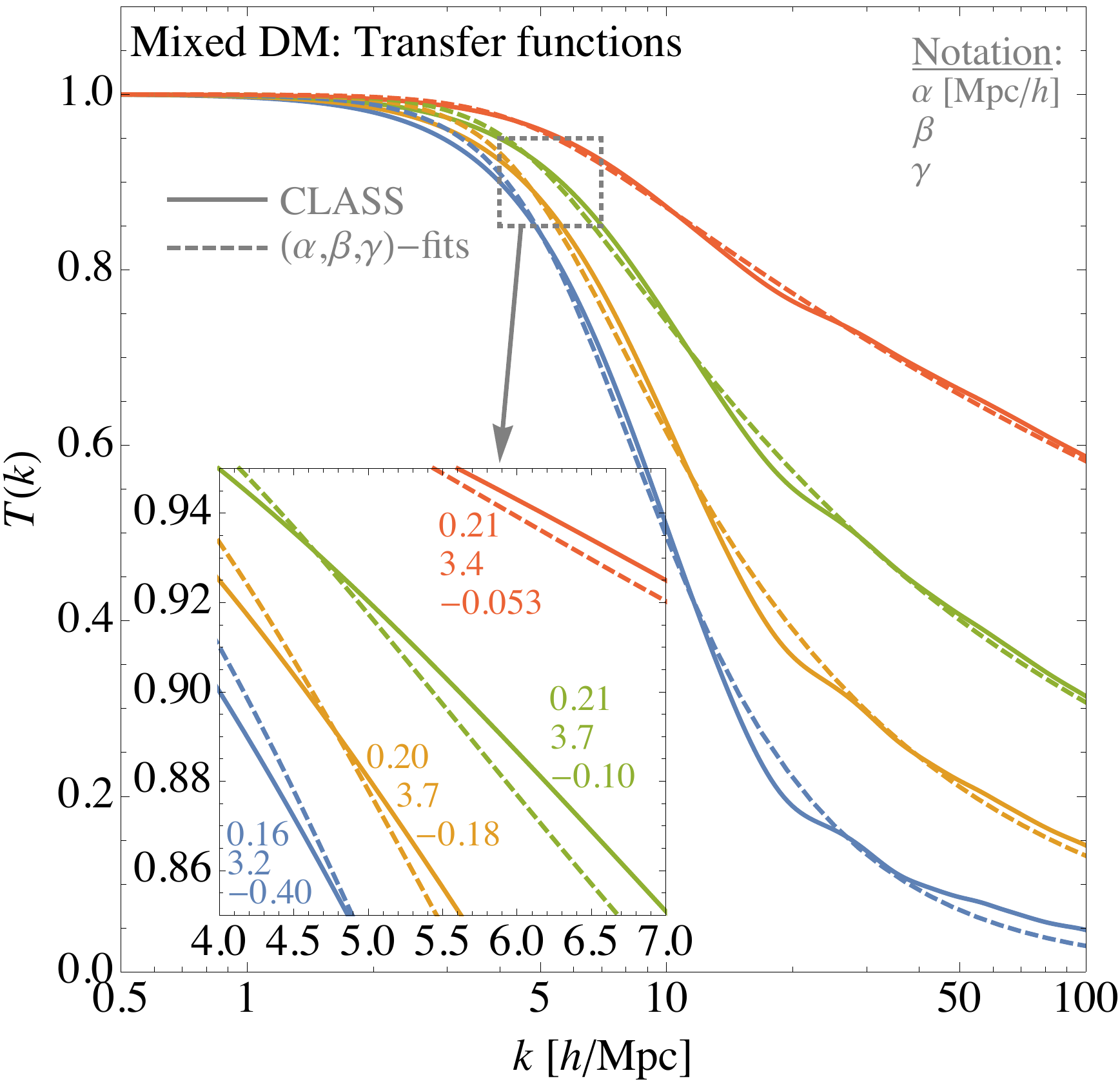}
  \end{tabular}
  \caption{\label{fig:Mixed-fits} Example power spectra (\emph{left}) and corresponding transfer functions (\emph{right}) for mixed DM, derived with {\tt CLASS}~\cite{Lesg:2011}. On the right panel, one can see that even the transfer functions featuring some type of plateau are fitted well by the parametrisation from Eq.~\eqref{eq:Tgen}.}
\end{figure}

Mixed DM is characterised by two parameters: the mass of the WDM component and the fraction $f$ of the warm to the total DM abundance, i.e., $f=\Omega_{\rm WDM}/\Omega_{\rm total}$, where $\Omega_{\rm total} = \Omega_{\rm WDM} + \Omega_{\rm CDM}$ denotes the total DM abundance in the universe. The fraction $f$ parametrises the example settings illustrated in Fig.~\ref{fig:Mixed-fits}, where we depict both the power spectra (\emph{left}) and the transfer functions (\emph{right}). For mixed DM, it had been pointed out in Ref.~\cite{Boyarsky:2008xj} that a non-zero plateau can be present in the transfer function for large $k$, corresponding to the remaining CDM component once the reduction of small scales by the warm component has died off. However, although our fitting function, Eq.~\eqref{eq:Tgen}, does not formally feature a plateau, it provides a very good fit to the majority of mixed DM cases, not only by eye but also when performing a goodness-of-fit test.

\subsection{\label{sec:models_Fuzzy}Fuzzy dark matter}

\looseness=-1 A conceptually different class of DM candidates that also affects the small scales of structure formation is the so-called Fuzzy DM.~\cite{Hu:2000ke,Marsh:2013ywa,Hui:2016ltb}. This type of DM consists of (initially) condensed scalar particles with tiny masses, $\sim 10^{-22}$~eV, such that their wave-like nature becomes relevant at astrophysical scales. These particles are assumed to have no self-interactions, quite like axions~{\cite{Hui:2016ltb}}, which could modify the picture if sufficiently strong~\cite{Rindler-Daller:2013zxa}. In the absence of such interactions, the class of Fuzzy DM models is conveniently described by a single parameter, namely the DM mass $m_{22} \equiv m_\psi / 10^{-22}$~eV, where $m_\psi$ denotes the actual physical particle mass. However, note that strong constraints exist on these scenarios, e.g.\ upper limits $m_{22} < 1.5$ from the kinematics of dwarf galaxies~\cite{Marsh:2015wka,Calabrese:2016hmp}, which were recently complemented by strong lower limits from the abundance of high-$z$ galaxies~\cite{Menci:2017nsr}, $m_{22} > 10$, superseeding earlier limits from their luminosity functions ($m_{22} > 1.2$~\cite{Schive:2015kza}) by nearly one order of magnitude. Even more recently, the Intergalactic Medium (IGM) have provided the tightest limits on the mass of $m_{22} > 20$ (2$\sigma$ C.L.) for a very conservative analysis of high redshift data~\cite{irsic17fuzzy} (while $m_{22} > 37.5$ is obtained for a less conservative scenario where some priors on the IGM thermal history are assumed).

\begin{figure}[t]
  \hspace{-0.5cm}
  \begin{tabular}{lr}
  \includegraphics[width=7.9cm]{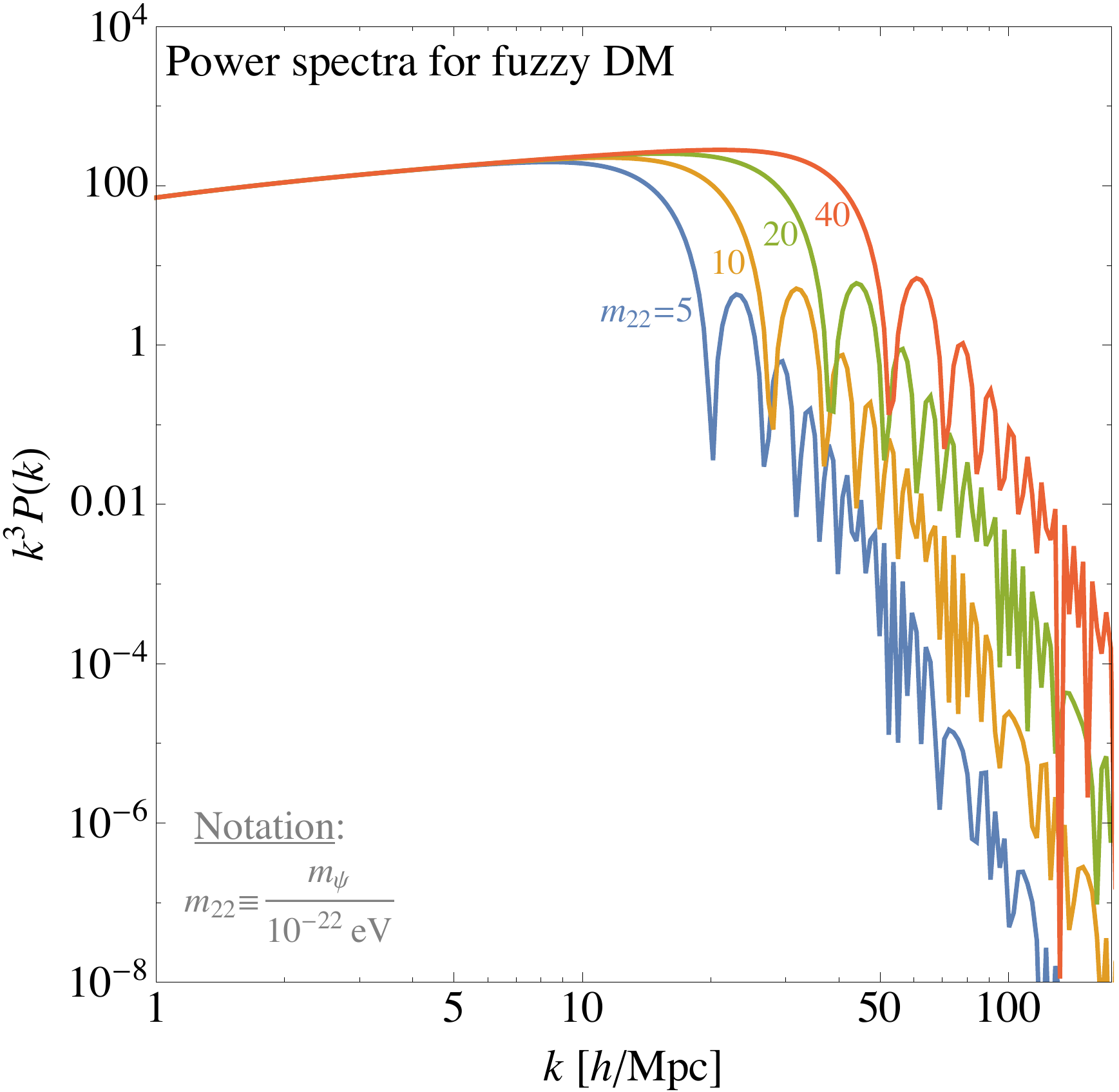} & \includegraphics[width=7.6cm]{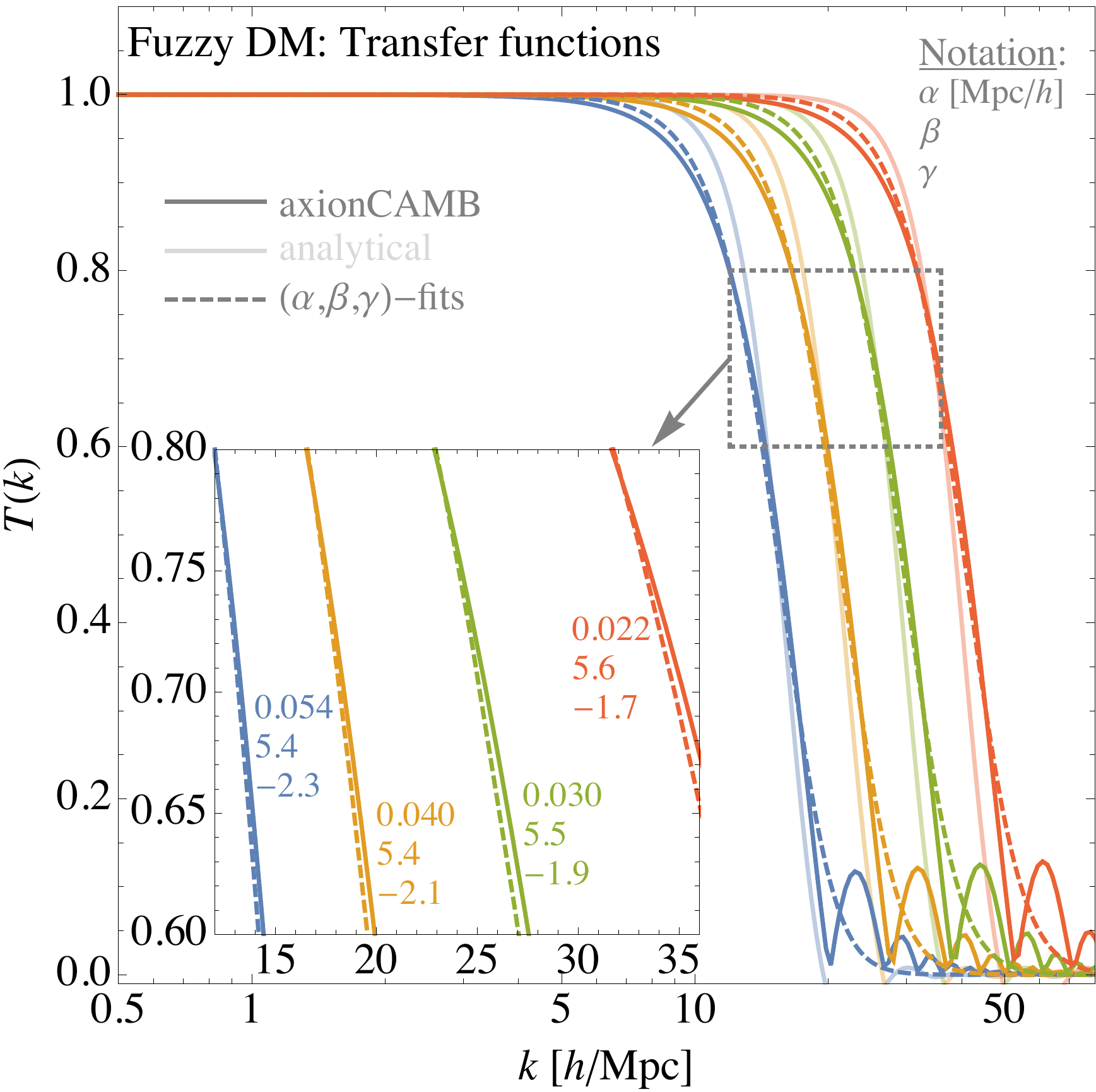}
  \end{tabular}
  \caption{\label{fig:Fuzzy-fits}{Example power spectra (\emph{left}) and corresponding transfer functions (\emph{right}) for fuzzy DM, derived with {\tt axionCAMB}~{\cite{hlozek15}} (on the right, we in addition show the analytical result from Ref.~\cite{Hu:2000ke}). On the right panel, one can see that the transfer functions are fitted very well by the parametrisation from Eq.~\eqref{eq:Tgen}.}}
\end{figure}

In Fig.~\ref{fig:Fuzzy-fits}, we show a few example power spectra (\emph{left}) and transfer functions (\emph{right}), associated with different values of $m_{22}$. The point that we want to illustrate is that, even though the known oscillations are present in the fuzzy DM power spectra, their transfer functions are still well described by our general parametrisation from Eq.~\eqref{eq:Tgen}, simply because the oscillations appear only at the smallest scales or, equivalently, at large values of $k$. To achieve a good fit, we have simply cut off the oscillations, which are very suppressed and therefore negligible for most applications, including the MW satellite counting and the Lyman-$\alpha$ forest methods.

\looseness=-1 Using this approximation, we can see that our parametrisation from Eq.~\eqref{eq:Tgen} in fact provides a very good fit to the fuzzy DM transfer functions, cf.\ right panel of Fig.~\ref{fig:Fuzzy-fits}. We will see in Sec.~\ref{sec:reality-check} how well this strategy truly works, when both the actual models and our fits are subjected to a reality test.

\subsection{\label{sec:models_ETHOS}Effective theory of structure formation (ETHOS)}

\begin{figure}[t]
  \hspace{-0.5cm}
  \begin{tabular}{lr}
  \includegraphics[width=7.9cm]{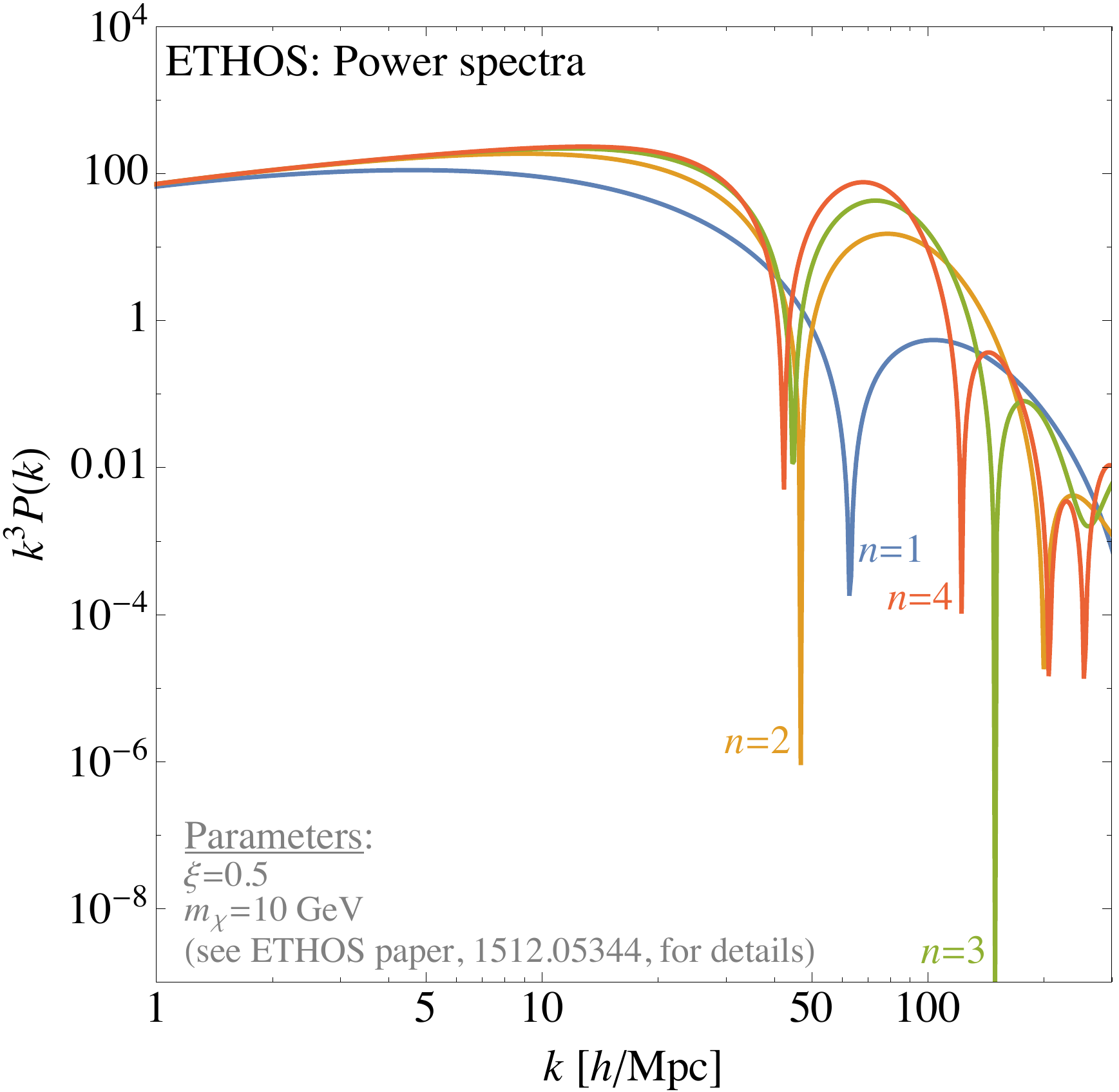} & \includegraphics[width=7.6cm]{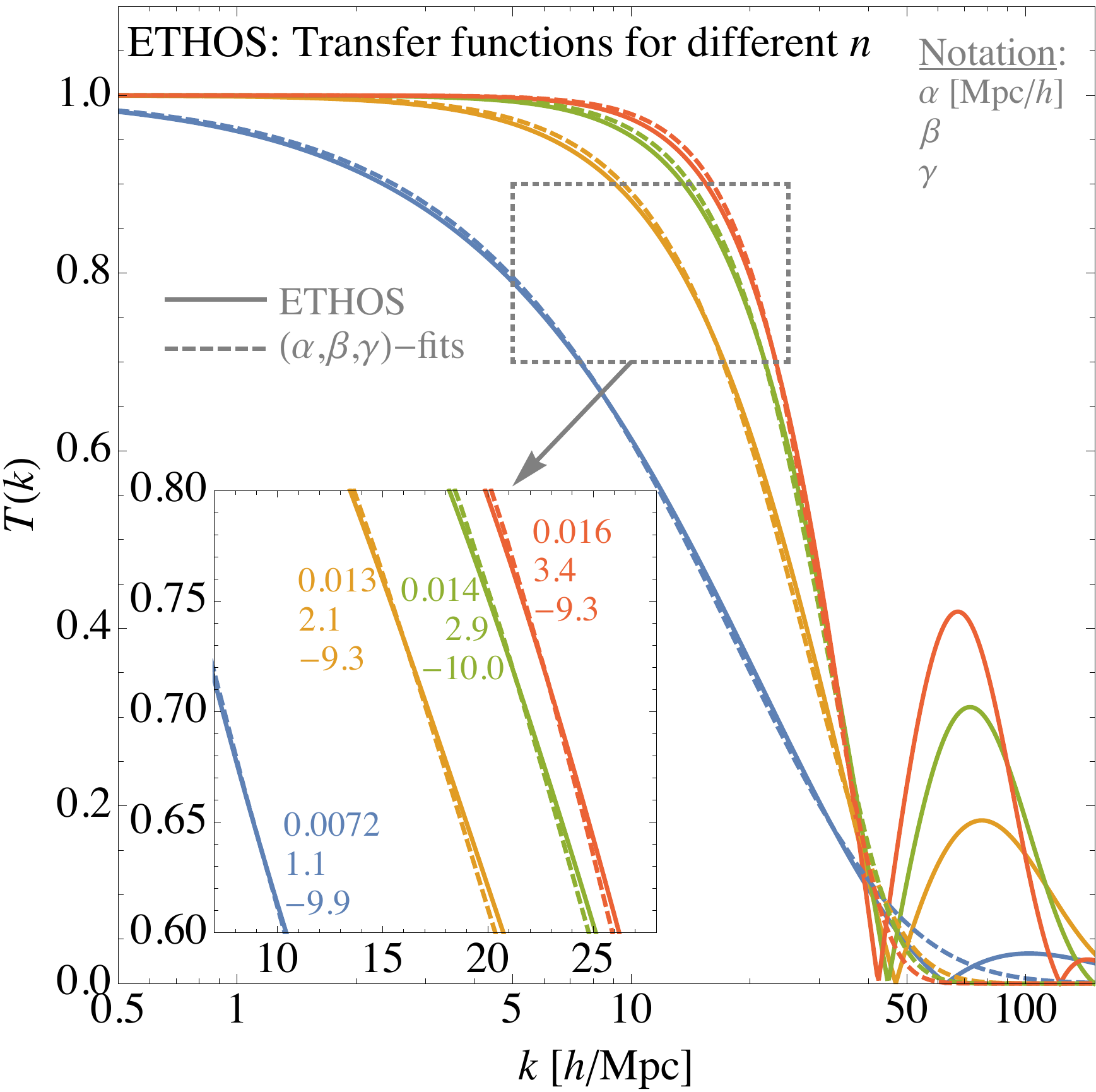}
  \end{tabular}
  \caption{\label{fig:ETHOS-fits}Power spectra (\emph{left}) and corresponding transfer functions (\emph{right}) for a few cases of self-interacting DM as derived with ETHOS, see Fig.~1a of Ref.~\cite{Cyr-Racine:2015ihg}. On the right panel, one can see that the transfer functions are fitted well by the parametrisation from Eq.~\eqref{eq:Tgen}, however, note that we have only fitted the part of $k$ left of the first ``oscillation''.}
\end{figure}

\looseness=-1 In order to further demonstrate the variability of our approach, we should compare it to that of ETHOS~\cite{Cyr-Racine:2015ihg,Vogelsberger:2015gpr}, which consists of an attempt to formulate an effective theory of cosmic structure formation, to map virtually any particle physics model to the constraints from astrophysics and cosmology. While to some extent this work is done in the same spirit as ours, we should point out the differences in our approach:
\begin{enumerate}

\item \emph{Langrangian-based vs.\ transfer function-based approach}:\\
\looseness=-1 While we are simply trying to parametrise the transfer function for nCDM, cf.\ Eq.~\eqref{eq:Tgen}, the approach of ETHOS is to start directly from the particle physics Lagrangian. While at first this seems like a clear disadvantage of our strategy, since a particle physicist would need to compute the matter power spectrum (e.g.\ using {\tt CLASS}~\cite{Lesg:2011}) before being able to apply our results to their model, we would like to point out that the mapping of DM models into the quantities relevant for cosmic structure formation is \emph{injective}. Put in simpler words, many models that look quite different from the particle physics point of view will yield precisely the same predictions for structure formation (this can be seen easily for thermal examples, e.g., by rescaling both temperature and DM mass). With our approach it is therefore possible to summarise most DM models in a much simpler framework. In fact, it may be a too big effort to start from any possible particle physics Lagrangian, when the key point for comparison lies in the transfer function.\\
\emph{We thus consider our approach to be the most economic.}

\item \emph{Validity for small scales, i.e., for large} $k$:\\
As we have pointed out e.g.\ in Sec.~\ref{sec:models_Fuzzy}, our approach cannot capture the very smallest scales, which are suppressed in the transfer function due to dividing by the CDM power spectrum. However, given that this is in any case the part of the spectrum with the smallest power, we do not expect big effects of this region, unless we find an observable that is truly sensitive to very small scales. The ones we are using (i.e., satellite counts and Lyman-$\alpha$) are not sensitive to that extreme region. For instance, as visible in the left panel of Fig.~\ref{fig:ETHOS-fits}, oscillations in the transfer functions seem very prominent when plotted in log-scales, although they are in fact unimportant for most aspects of structure formation.\\
\emph{We thus consider our approach to be safe as long as the regions for (very) large $k$ play no role.}

\item \emph{Model coverage}:\\
By construction, our approach of fitting the transfer function is much simpler and therefore less versatile than a method starting from the particle Lagrangian. However, we want to point that the original fitting function of Eq.~\ref{eq:Tgen} can be easily generalised if necessary. For example, a plateau in the transfer function can be described by adding one more parameter $\delta$, i.e.
\begin{equation}
 T(k) = [ 1 + (\alpha k)^{\beta} ]^{\gamma} \to T(k) = [ 1 + (\alpha k)^{\beta} ]^{\gamma} + \delta.
 \label{eq:Tgen_extended}
\end{equation}
In principle, oscillationary patterns could also be included, for example by simply adding a cosine function of the form
\begin{equation}
 T(k) = [ 1 + (\alpha k)^{\beta} ]^{\gamma} \cdot \cos^2 (\delta k)
\label{eq:Tgen_extended2}
\end{equation}
where $\delta$ is an additional free parameter.\\
\emph{We thus consider our approach to be easily extendable to basically cover the same range of models as ETHOS does.}

\end{enumerate}

Given that, we can try to fit some of the transfer functions obtained by ETHOS, and for this purpose we take the ones given in Fig.~1a of Ref.~\cite{Cyr-Racine:2015ihg} as example (see that reference for the details on the data chosen).\footnote{Note that Ref.~\cite{Cyr-Racine:2015ihg} uses an alternative definition of the transfer function of $T_{\rm ETHOS}(k) \equiv P(k)_{\rm{nCDM}}/P(k)_{\rm{CDM}}$, instead of the more common definition of $T^2(k) \equiv P(k)_{\rm{nCDM}}/P(k)_{\rm{CDM}}$, which is the one we are using. This difference in definitions is what creates the seeming difference between the right panel of our Fig.~\ref{fig:ETHOS-fits} and Fig.~1a of Ref.~\cite{Cyr-Racine:2015ihg}.} This will allow us to investigate whether our approach yields a similar result as the more detailed ETHOS analysis. We have again fitted the transfer functions with Eq.~\eqref{eq:Tgen}, however, note that this time we have only fitted the part for small $k$, i.e., left of the first ``oscillation''. In Sec.~\ref{sec:reality-check} we will show that regarding the number of Milky-Way satellites as well as the power spectrum from the Lyman-$\alpha$ forest, there is hardly any difference between our simplifying fit without oscillations and the full transfer function from ETHOS.

\section{\label{sec:results}Results from ${\boldsymbol{N}}$-body simulations}

In order to perform cosmological $N$-body simulations and to put constraints on the nCDM models listed in Tab.~\ref{tab:params}, we have modified the numerical code {\tt 2LPT}~\cite{Crocce:2006ve} -- which generates initial conditions for running cosmological simulations -- by implementing the new, general transfer function: now the code takes as inputs $\{\alpha,\beta,\gamma\}$ instead of the thermal WDM mass, and it computes the corresponding $T(k)$ from Eq.~\eqref{eq:Tgen}.

In the right panel of Fig.~\ref{fig:Tk} we plot the matter power spectra associated with the combinations of $\{\alpha,\beta,\gamma\}$ listed in Tab.~\ref{tab:params}, computed at redshift $z=99$ by using the modified version of {\tt 2LPT}, plus the matter power spectrum associated to a purely CDM universe (the dashed line in the plot). The dotted line refers to the matter power spectrum computed, in linear theory, with the numerical Boltzmann solver {\tt CLASS}~\cite{Lesg:2011}  in a $\Lambda$CDM universe at $z=99$ as well. All the relevant cosmological parameters are set to their Planck values~\cite{Adam:2015rua}: $\Omega_{m}=0.317$, $\Omega_{\Lambda}=0.683$, $h=0.67$, $\sigma_8=0.8$.

We have used these snapshots as initial conditions for running 55+1 DM-only simulations (512$^3$ particles in a 20 Mpc$/h$ box) with the {\tt GADGET-3} code, a modified version of the publicly available {\tt GADGET-2} code~\cite{Springel:2005mi,Springel:2000yr}. On top of these simulations, we have used a Friends-of-Friends (FoF) algorithm~\cite{Davis:1985rj} with the standard linking length $b=0.2$, in order to extract the DM halos. We also run the {\tt SUBFIND} code~\cite{Springel:2000qu} for searching for the substructures bound to each main FoF group.

The suite of simulations used here has the following purposes: $i)$ present a first assessment of non-linearities for the models discussed; $ii)$ address quantitatively how the mass functions based on linear theory predictions compare to the actual results of the $N$-body simulations; $iii)$ assess whether non-linearities in the matter power spectrum could affect the conclusions regarding the derived constraints from our approximate Lyman-$\alpha$ method. In the following, we will see that the constraints given are primarily based on the linear theory and the simulations are mainly used as a first cross-check that the results are indeed not altered by non-linearities and/or by a more accurate modeling of either the number of MW satellites or the IGM structures.

\subsection{\label{subsec:mpk}Non-linear matter power spectra}

\begin{figure}[t]
  \begin{tabular}{lr}
  \includegraphics[page=1,width=7.7cm]{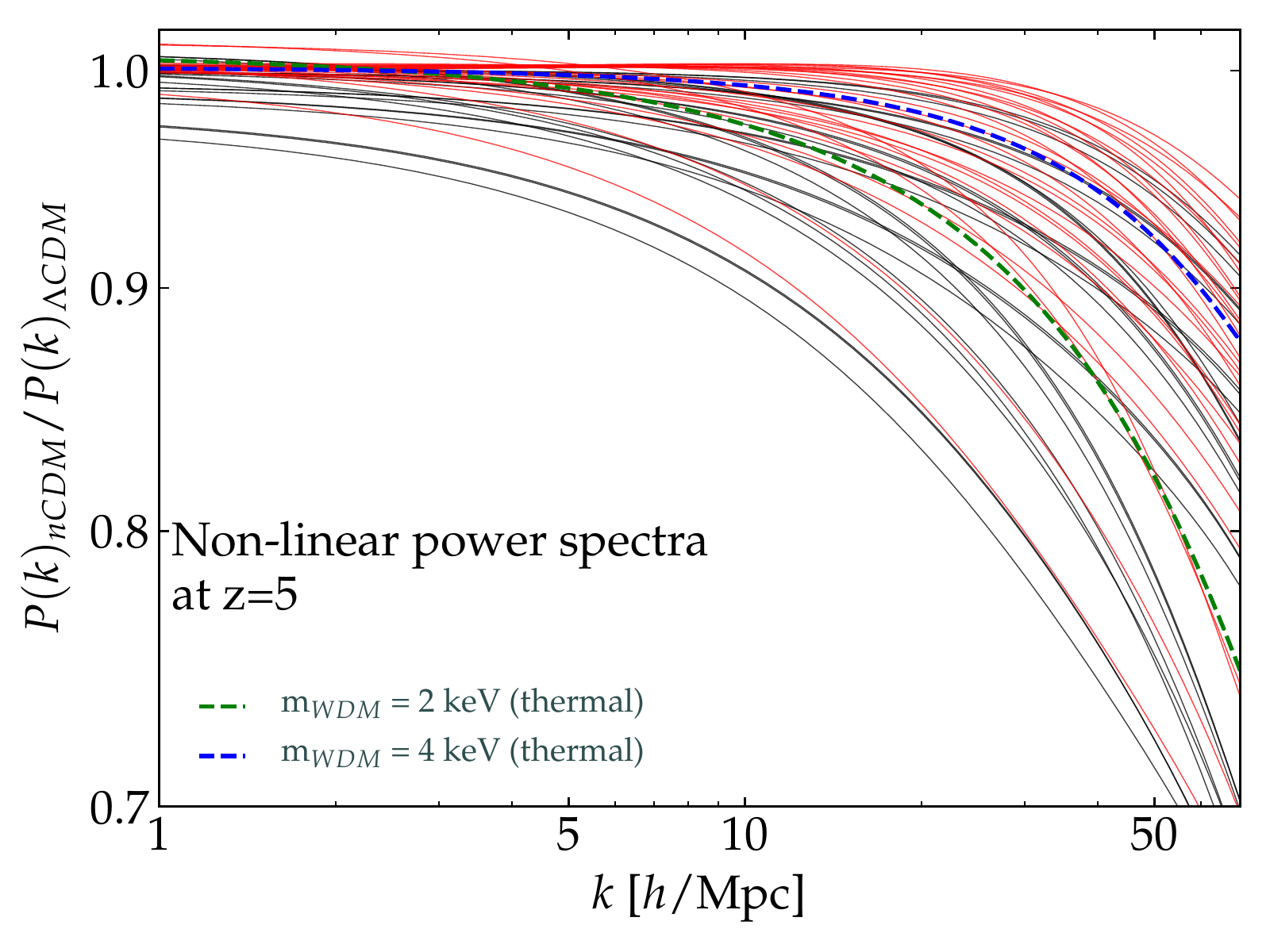} & \includegraphics[page=2,width=7.7cm]{./new_pics/NLpk_ALL_z2}
  \end{tabular}
  \caption{\label{fig:mpk}Ratios of the nCDM non-linear matter power spectra with respect to the $\Lambda$CDM power spectrum, at redshifts $z=5$ and $z=2$.
  The black lines correspond to the first 27 models listed in Table~\ref{tab:params}, while the red lines correspond to the additional 28 models. The green and blue dashed lines refer to thermal WDM models with masses of 2 and 4~keV, respectively.}
\end{figure}

In Fig.~\ref{fig:mpk} we plot the ratios between the nCDM non-linear matter power spectra with respect to the $\Lambda$CDM power spectrum.
We show the power spectra of all nCDM models at redshift 5 (left panel) and redshift 2 (right panel). Additionally, we illustrate the thermal WDM cases with 2 and 4 keV, which are in agreement with a similar study made by Ref.~\cite{viel12}.
The differences between the models gradually decrease when going to smaller redshifts. Below redshift 2, the small-scale power enhancement from the non-linear structure evolution starts to dominate the signal from different nCDM models, resulting in a progressive shift of the corresponding half-mode scales towards larger values of $k$. This evolution is the reason why the Lyman-alpha data from the highest redshift bins provide the strongest limits on the nature of DM.

\subsection{\label{subsec:hmf}Halo mass functions}

\begin{figure}[b]
  \begin{tabular}{lr}
  \includegraphics[page=1,width=7.7cm]{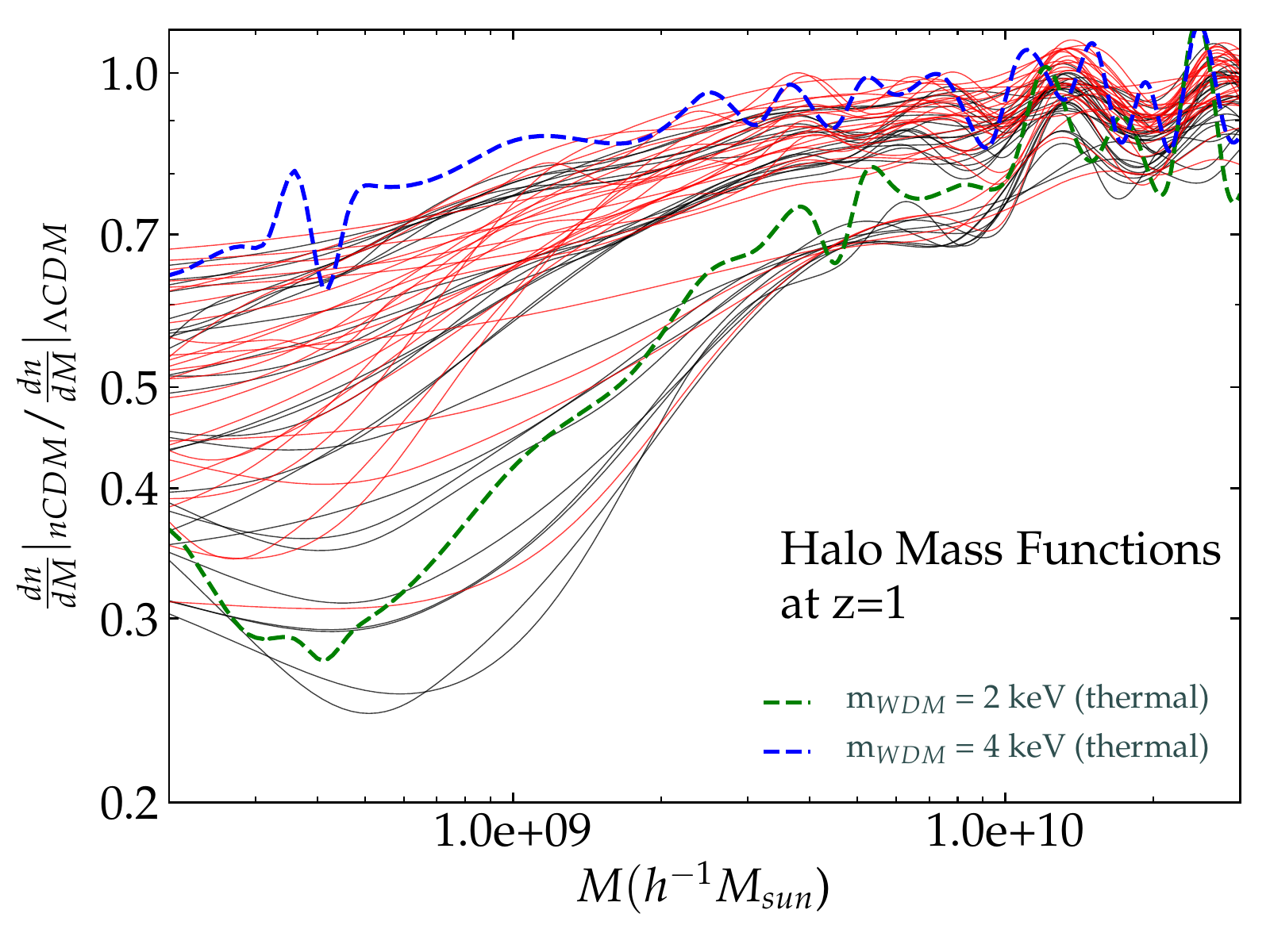} & \includegraphics[page=2,width=7.7cm]{./new_pics/MF_ALL}
  \end{tabular}
  \caption{\label{fig:hmf}Ratios of the nCDM halo mass functions with respect to the $\Lambda$CDM halo mass function, at redshifts $z=1$ and $z=0$. The black lines correspond to the first 27 models listed in Table~\ref{tab:params}, while the red lines correspond to the additional 28 models. The upturn at low masses is due to artificial clumping, while the oscillation pattern at large masses is due to the cosmic variance. The green and blue dashed lines refer to thermal WDM models with masses of 2 and 4~keV, respectively.}
\end{figure}

\looseness=-1 In Fig.~\ref{fig:hmf} we plot the ratios between the nCDM halo mass functions with respect to the $\Lambda$CDM halo mass function, at redshift 1 ({\emph{left}}) and 0 ({\emph{right}}). Note that all the nCDM models predict a lower abundance of halos with masses between $10^8$ and $10^9$ $M_\odot/h$: for some of the $\{\alpha,\beta,\gamma\}$-triplets the number of halos included in this mass range is expected to be 80$\%$ smaller with respect to $\Lambda$CDM predictions.

The visible upturn at low masses is not physical, but a consequence of the phenomenon of {\emph{artificial clumping}}, due the limited resolution of our simulations, which affects models with suppressed initial power spectra~\cite{Angulo:2013sza,Schneider:2013ria,Schneider:2014rda}. This is why it is crucial to have good theoretical predictions for the low-mass behaviour of the halo (and subhalo) mass functions. This issue, essential for the analyses presented in Sec.~\ref{subsec:satellite}, is discussed in more detail in the Appendix~\ref{ap:numconv}. The oscillatory pattern which characterises the region corresponding to masses $\gtrsim 10^{10}~M_{\odot}/h$ is due to the cosmic variance, since the size of our simulated box does not allow to have enough large halos to have statistically fully meaningful results.

\section{\label{sec:obs}Comparison with observations}

In this section we present the first constraints on $\{\alpha, \beta, \gamma\}$ from structure formation data, based on linear theory. We have constrained the parameters by using MW satellite counts and Lyman-$\alpha$ forest data, which constitute two powerful independent methods for testing the ``non-coldness'' of DM. Satellite counts rely on the simple fact that any nCDM model must predict a number of substructures within the MW virial radius not smaller than the actual number of MW satellites that we observe. \looseness=-1 The Lyman-$\alpha$ forest data analysis instead provides information on the matter power spectrum along our line of sight, down to scales corresponding to  $k \sim 10~h/$Mpc, currently giving the most stringent limits on the masses of thermal WDM candidates,  which is why they can also be expected to be strong for non-thermal cases.

\subsection{\label{subsec:satellite}Constraints from Milky Way satellite counts}

\begin{table}[b]
\begin{center}
  \begin{tabular}{|c|c|}
  \hline
				   &  $N_{\rm{sub}}$		\\ \hline
\scriptsize{nCDM1}	   		&  39 		      \\ 
{\bf \scriptsize{nCDM2}}	   &  {\bf 78} 		      \\ 
{\bf \scriptsize{nCDM3}}	   &  {\bf 105} 		      \\ 
\scriptsize{nCDM4}	   &  25 		     \\ 
{\bf \scriptsize{nCDM5}}	   &  {\bf 68} 		      \\ 
{\bf \scriptsize{nCDM6}}	   &  {\bf 104} 		      \\ 
\scriptsize{nCDM7}	   &  18 		     \\ 
\scriptsize{nCDM8}	   &  59 		     \\ 
{\bf \scriptsize{nCDM9}}	   &  {\bf 103} 		     \\
\scriptsize{nCDM10}      &  41 		     \\ 
{\bf \scriptsize{nCDM11}}	   &  {\bf 78} 		      \\ \hline 
  \end{tabular}                       
  \begin{tabular}{|c|c|}
  \hline
			     &  $N_{\rm{sub}}$ 	\\ \hline

{\bf \scriptsize{nCDM12}}	   &  {\bf 105} 		      \\ 
\scriptsize{nCDM13}	   &  27 		     \\ 
{\bf \scriptsize{nCDM14}}	   &  {\bf 69} 		     \\ 
{\bf \scriptsize{nCDM15}}	   &  {\bf 104} 		      \\ 
\scriptsize{nCDM16}	   &  19 		     \\ 
\scriptsize{nCDM17}	   &  60 		     \\ 
{\bf \scriptsize{nCDM18}}	   &  {\bf 103} 		     \\ 
\scriptsize{nCDM19} 	   &  52 		   \\ 
{\bf \scriptsize{nCDM20}}  &  {\bf 83}	   \\ 
{\bf \scriptsize{nCDM21}}  &  {\bf 106}	    \\ 
\scriptsize{nCDM22}  &  38	   \\ \hline
  \end{tabular}
  \begin{tabular}{|c|c|}
  \hline
			     &  $N_{\rm{sub}}$ 	\\ \hline

{\bf \scriptsize{nCDM23}}  &  {\bf 75}	   \\ 
{\bf \scriptsize{nCDM24}}  &  {\bf 105}	   \\ 
\scriptsize{nCDM25}  &  28	   \\ 
{\bf \scriptsize{nCDM26}}  &  {\bf 68}	   \\ 
{\bf \scriptsize{nCDM27}}  &  {\bf 104}	   \\ 
{\bf \red\scriptsize{nCDM28}}  &  {\bf \red76} 	  \\ 
{\bf \red\scriptsize{nCDM29}}  &  {\bf \red89} 	    \\
{\bf \red\scriptsize{nCDM30}}  &  {\bf \red69} 	  \\ 
{\bf \red\scriptsize{nCDM31}}  &  {\bf \red85} 	  \\ 
\red\scriptsize{nCDM32}  & \red49 	  \\ 
{\bf \red\scriptsize{nCDM33}}  &  {\bf \red68} 	  \\ \hline
 \end{tabular}                      
  \begin{tabular}{|c|c|}
  \hline
			     &  $N_{\rm{sub}}$ 	\\ \hline

\red\scriptsize{nCDM34}  &  \red42 	  \\ 
\red\scriptsize{nCDM35}  &  \red57 	  \\
\red\scriptsize{nCDM36}  &  \red52	  \\
{\bf \red\scriptsize{nCDM37}}  &  {\bf \red71}	 \\
{\bf \red\scriptsize{nCDM38}}  &  {\bf \red81}	  \\     
{\bf \red\scriptsize{nCDM39}}  &  {\bf \red90}     \\
{\bf \red\scriptsize{nCDM40}}  &  {\bf \red106}      \\
{\bf \red\scriptsize{nCDM41}}  &  {\bf \red106}      \\
{\bf \red\scriptsize{nCDM42}}  &  {\bf \red86}      \\
{\bf \red\scriptsize{nCDM43}}  &  {\bf \red93}     \\
{\bf \red\scriptsize{nCDM44}}  &  {\bf \red95}      \\ \hline
  \end{tabular}
  \begin{tabular}{|c|c|}
  \hline
			     &  $N_{\rm{sub}}$  	\\ \hline

{\bf \red\scriptsize{nCDM45}}  &  {\bf \red99}      \\
\red\scriptsize{nCDM46}  &  \red26      \\
\red\scriptsize{nCDM47}  &  \red45     \\
{\bf \red\scriptsize{nCDM48}}  &  {\bf \red63}      \\
{\bf \red\scriptsize{nCDM49}}  &  {\bf \red76}      \\
{\bf \red\scriptsize{nCDM50}}  &  {\bf \red105}      \\
{\bf \red\scriptsize{nCDM51}}  &  {\bf \red106}      \\
{\bf \red\scriptsize{nCDM52}}  &  {\bf \red71}      \\
{\bf \red\scriptsize{nCDM53}}  &  {\bf \red82}      \\
{\bf \red\scriptsize{nCDM54}}  &  {\bf \red86}      \\
{\bf \red\scriptsize{nCDM55}}  &  {\bf \red93}      \\ \hline  
\end{tabular}                   
\caption{\label{tab:subhalos}{Number of subhalos (with mass $M_{\rm sub} \geq 10^8~M_\odot/h$) within the virial radius of a halo with mass $M_{\rm halo} = 1.7 \cdot 10^{12}~M_\odot/h$. Each of the 55 models corresponds to a different $\{\alpha, \beta, \gamma\}$-combination, according to Tab.~\ref{tab:params}. Models that predict a number of subhalos consistent with observations (i.e., predicting at least as many subhalos as the number of observed MW satellites, $N_{\rm sat} = 63$) are written in bold-face, while those not surviving the reality check are not bold-faced.}}
\end{center}
\end{table}
\begin{figure}[t]
  \centering
   \subfigure{\includegraphics[width=7.5cm]{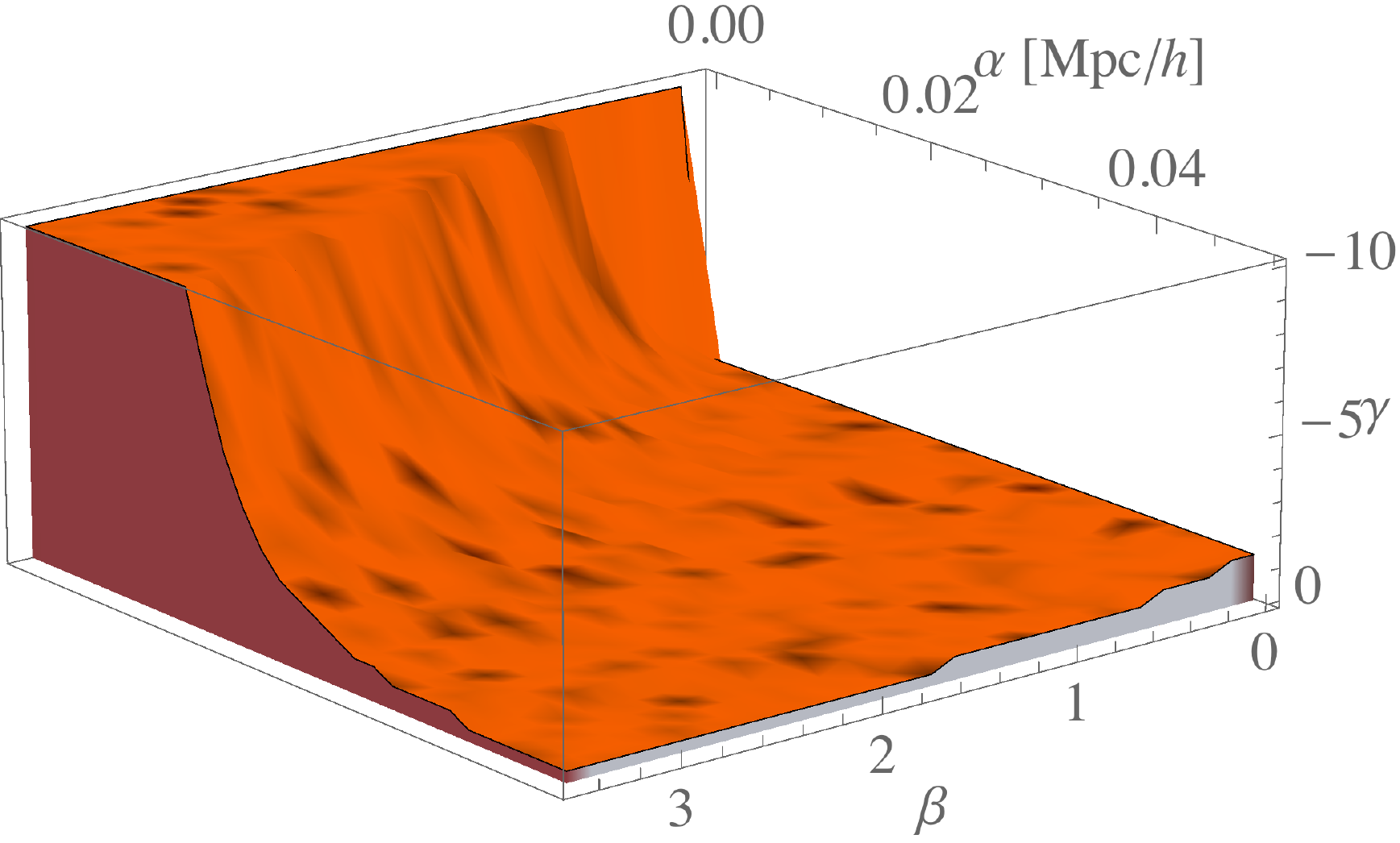}}
  \subfigure{\includegraphics[width=7.5cm]{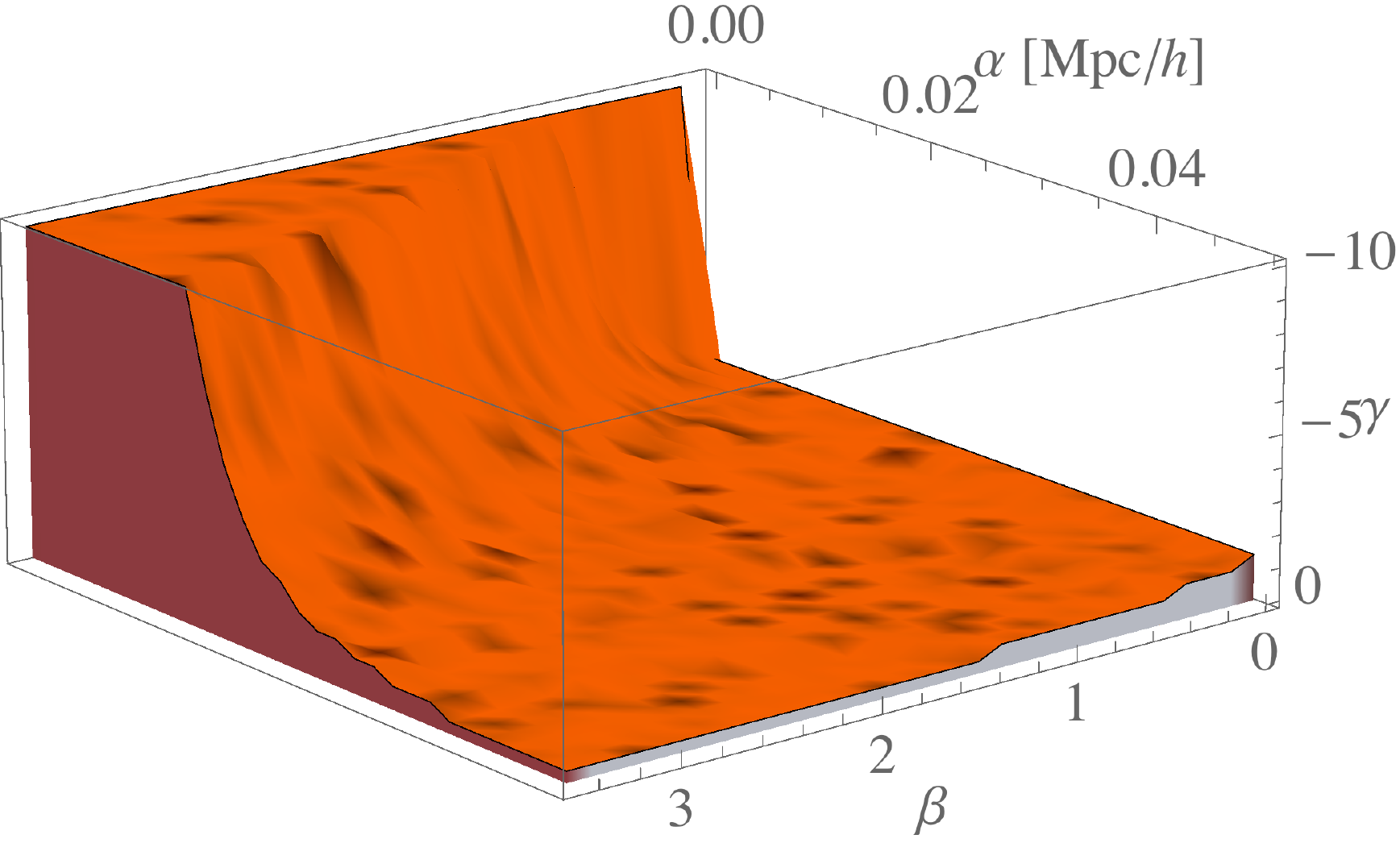}}
  \caption{\label{fig:sub3d} The 3-dimensional orange contour in the $\{\alpha, \beta, \gamma\}$-space represents the upper bound (i.e., with the largest modulus for $\gamma$) on the region of the parameter space which is in agreement with MW satellite counts. The left panel refers to the conservative analysis ($N_{\rm sub} \geq 57$), whereas the right panel refer to the non-conservative case ($N_{\rm sub} \geq 63$).}
\end{figure}

Assuming the standard $\Lambda$CDM model, cosmological $N$-body simulations predict too many dwarf galaxies within the MW virial radius, with respect to observations. Therefore, MW satellite counts represent a very useful tool for constraining DM properties (see, e.g.,~Refs.~\cite{2016arXiv161207834J, Vegetti:2009gw, Vegetti:2012mc, Vegetti:2014lqa}). We can estimate the number of MW satellites following the approach of~\cite{Polisensky:2010rw, Schneider:2016uqi}, i.e., by multiplying the 15 ultra-faint satellites from SDSS by a factor 3.5, in order to account for the limited sky coverage of the survey, and by finally summing the 11 MW classical satellites. We obtain $N_{\rm sat} = 63$ as an estimate of the number of observed satellites within the MW viral radius.

We can now compare $N_{\rm sat}$ with the number of subhalos $N_{\rm sub}$ predicted by our models, in order to extract some constraints on $\{\alpha, \beta, \gamma\}$. According to~\cite{Schneider:2016uqi,Schneider:2014rda}, we can use the following formula for estimating the number of substructures for a given model:
\begin{equation}\label{eq:subMF}
 \frac{{\rm d}N}{{\rm d}M_{\rm sub}} = \frac{1}{44.5}\frac{1}{6\pi^2}\frac{M_{\rm halo}}
 {M^2_{\rm sub}}\frac{P(1/R_{\rm sub})}{R^3_{\rm sub}\sqrt{2\pi(S_{\rm sub}-S_{\rm halo})}},
\end{equation}
where $M_{\rm sub}$ and $S_{\rm sub}$ are the mass and the variance of a given subhalo, $M_{\rm halo}$ and $S_{\rm halo}$ are the mass and the variance of the main halo, defined as follows:
\begin{equation}\label{eq:mass_var}
 S_i = \frac{1}{2\pi^2} \int\limits_0^{1/R_i} {\rm d}k\ k^2P(k); ~~~~~ M_i = \frac{4\pi}{3}\Omega_m\rho_c(cR_i)^3; ~~~~~ c=2.5;
\end{equation}
with $P(k)$ being the linear power spectrum of a given model, computed at redshift $z=0$. In Appendix~\ref{ap:numconv} we explicitly show that the mass functions for the grid of simulations performed in the present work are in
good agreement with the theoretical mass  function formalism outlined above.

\looseness=-1 Under the assumption of a MW halo mass $M_{\rm halo} = 1.7 \cdot 10^{12}~M_\odot/h$~\cite{Lovell:2013ola} and by considering subhalos with masses $M_{\rm sub} \geq 10^8~M_\odot/h$, we can obtain the number of subhalos $N_{\rm sub}$ predicted by our models, by simply integrating Eq.~\eqref{eq:subMF}. The results are reported in Tab.~\ref{tab:subhalos}, where the models highlighted in bold-face are those in agreement with the number of observed satellites, i.e.\ with $N_{\rm sub} \geq 63$.

\looseness=-1 In Fig.~\ref{fig:sub3d} we show a 3-dimensional contour plot in the $\{\alpha, \beta, \gamma\}$-space, where each triplet is associated with a different model: the orange contour represents the upper bound on the region of the parameter space which is in agreement with MW satellite counts, according to the method that we have just outlined. Hence, in the right panel of Fig.~\ref{fig:sub3d}, all the $\{\alpha, \beta, \gamma\}$-combinations which sample the orange volume correspond to models that predict a number of substructures at least equal to $N_{\rm sat}=63$. In the left panel, instead, we plot the allowed volume of the parameter space whether we require the number of subhalos predicted by each nCDM model to be equal or larger with respect to a more conservative estimate for the number of MW satellites, $N_{\rm sat}=57$. This number has been chosen in order to account for a ten percent sampling variance in the number of satellites at a given MW halo mass.

By marginalising over $\beta$ and $\gamma$ we obtain the following limits on $\alpha$:
\begin{equation}\label{eq:alfalimit_sub}
\begin{matrix}
 \alpha \leq 0.067~{\rm{Mpc}}/h~~(\rm{95\% ~C.L.}) & & & & & & & & & \text{requiring}~N_{\rm sub} \geq 63,\hfill\hfill\hfill\hfill\hfill\\
 \alpha \leq 0.061~{\rm{Mpc}}/h~~(\rm{95\% ~C.L.}) & & & & & & & & & \text{requiring}~N_{\rm sub} \geq 57,
 \end{matrix}
\end{equation}
which would correspond, in the old one-to-one parametrisation, to a thermal WDM particle with mass $m_{\rm{WDM}} \approx 2 ~\rm{keV}$ (see Eq.~\eqref{eq:Viel} and Eq.~\eqref{eq:alphaold}). These limits are less constraining than the latest constraints from structure formation data: as expected, modeling the power suppression with three free parameters weakens the constraints on $k_{1/2}$. Within our general approach, due to the dependence of $\alpha$ on $\beta$ and $\gamma$, even lighter DM candidates may provide suppressed power spectra in agreement with MW satellite counts.

Looking at Eq.~\eqref{eq:alfalimit_sub}, it may seem surprising that the constraint on $\alpha$ strengthens when we use a weaker rejection criterion for the nCDM models, i.e.~a smaller value for $N_{\rm sat}$. The reason is that, by relaxing the limit on the number of substructures, we accept a larger number of $\{\alpha, \beta, \gamma\}$-triplets characterised  by very small values of $\alpha$. This is a straightforward consequence of the shape of the volume of the parameter space shown in Fig.~\ref{fig:sub3d}. By accepting all those models which predict $57 \leq N_{\rm sub} < 63$, we are slightly shifting towards higher values of $\alpha$ the whole orange 3-dimensional surface in the plot. As one can see by comparing the two panels of Fig.~\ref{fig:sub3d}, due to the geometry of the orange contour, this shift mainly concerns $\{\alpha, \beta, \gamma\}$-combinations with $\alpha$ very close to zero. Therefore, by imposing a smaller value for $N_{\rm sat}$ and marginalising over $\beta$ and $\gamma$, small values of $\alpha$ increase their contribution to its probability distribution with respect to the high-value tail of the distribution, which instead is only minimally affected by the choice of a more conservative value for $N_{\rm sat}$. We are therefore pushing the peak of the probability distribution of $\alpha$ towards zero, yielding to a stronger upper bound on it. However, at this approximate level of analysis, the difference between the two bounds reported in Eq.~\eqref{eq:alfalimit_sub} is practically negligible, as visible in Fig.~\ref{fig:sub3d}.


\subsection{\label{subsec:lyman}Constraints from the Lyman-$\alpha$ forest data}

\begin{table}[t]
\begin{center}
  \begin{tabular}{|c|c|}
  \hline
				   &  $\delta A$		\\ \hline
\scriptsize{nCDM1}	   &  0.61	      \\ 
\scriptsize{nCDM2}	   &  0.45	      \\ 
{\bf \scriptsize{nCDM3}}	   &  {\bf 0.34}	      \\ 
\scriptsize{nCDM4}	   &  0.63	     \\ 
\scriptsize{nCDM5}	   &  0.45	      \\ 
{\bf \scriptsize{nCDM6}}	   &  {\bf 0.32}	      \\ 
\scriptsize{nCDM7}	   &  0.64	     \\ 
\scriptsize{nCDM8}	   &  0.45	     \\ 
{\bf \scriptsize{nCDM9}}	   &  {\bf 0.31}	     \\
\scriptsize{nCDM10}    	   &  0.61	     \\ 
\scriptsize{nCDM11}	   &  0.45	      \\ \hline
\end{tabular}
\begin{tabular}{|c|c|}
  \hline
				   &  $\delta A$		\\ \hline
{\bf \scriptsize{nCDM12}}	   &  {\bf 0.34}	      \\ 
\scriptsize{nCDM13}	   &  0.63	     \\ 
\scriptsize{nCDM14}	   &  0.45	     \\ 
{\bf \scriptsize{nCDM15}}	   &  {\bf 0.32}	      \\ 
\scriptsize{nCDM16}	   &  0.64	     \\ 
\scriptsize{nCDM17}	   &  0.45	     \\ 
{\bf \scriptsize{nCDM18}}	   &  {\bf 0.31}	     \\ 
\scriptsize{nCDM19} 	   &  0.59	   \\ 
\scriptsize{nCDM20}  &   0.44  \\ 
{\bf \scriptsize{nCDM21}}  &   {\bf 0.34}   \\ 
\scriptsize{nCDM22}  &   0.61  \\ \hline
\end{tabular}
  \begin{tabular}{|c|c|}
  \hline
			     &  $\delta A$ 	\\ \hline
\scriptsize{nCDM23}  &   0.44  \\ 
{\bf \scriptsize{nCDM24}}  &   {\bf 0.32}  \\ 
\scriptsize{nCDM25}  &   0.62  \\ 
\scriptsize{nCDM26}  &   0.44  \\ 
{\bf \scriptsize{nCDM27}}  &   {\bf 0.31}  \\ 
\red\scriptsize{nCDM28}  &   \red0.42 \\ 
{\bf \red\scriptsize{nCDM29}}  &   {\bf \red0.37}   \\
\red\scriptsize{nCDM30}  &   \red0.42 \\ 
{\bf \red\scriptsize{nCDM31}}  &   {\bf \red0.37} \\ 
\red\scriptsize{nCDM32}  &   \red0.40 \\ 
{\bf \red\scriptsize{nCDM33}}  &   {\bf \red0.34} \\  \hline
\end{tabular}
  \begin{tabular}{|c|c|}
  \hline
			     &  $\delta A$ 	\\ \hline
\red\scriptsize{nCDM34}  &   \red0.39 \\ 
{\bf \red\scriptsize{nCDM35}}  &   {\bf \red0.33} \\
\red\scriptsize{nCDM36}  &   \red0.57 \\
\red\scriptsize{nCDM37}  &   \red0.54 \\
\red\scriptsize{nCDM38}  &   \red0.42 \\          
\red\scriptsize{nCDM39}  &  \red0.42    \\
{\bf \red\scriptsize{nCDM40}}  &  {\bf \red0.31}      \\
{\bf \red\scriptsize{nCDM41}}  &  {\bf \red0.33}      \\
\red\scriptsize{nCDM42}  &  \red0.40     \\
\red\scriptsize{nCDM43}  &  \red0.40    \\
{\bf \red\scriptsize{nCDM44}}  &  {\bf \red0.36}     \\ \hline
  \end{tabular}
  \begin{tabular}{|c|c|}
  \hline
			     &  $\delta A$  	\\ \hline

{\bf \red\scriptsize{nCDM45}}  &  {\bf \red0.37}     \\
\red\scriptsize{nCDM46}  &  \red0.61     \\
\red\scriptsize{nCDM47}  &  \red0.57    \\
\red\scriptsize{nCDM48}  &  \red0.41     \\
\red\scriptsize{nCDM49}  &  \red0.40     \\
{\bf \red\scriptsize{nCDM50}}  &  {\bf \red0.27}      \\
{\bf \red\scriptsize{nCDM51}}  &  {\bf \red0.27}      \\
{\bf \red\scriptsize{nCDM52}}  &  {\bf \red0.38}     \\
{\bf \red\scriptsize{nCDM53}}  &  {\bf \red0.37}     \\
{\bf \red\scriptsize{nCDM54}}  &  {\bf \red0.33}     \\
{\bf \red\scriptsize{nCDM55}}  &  {\bf \red0.33}     \\ \hline  
\end{tabular}
\caption{\label{tab:lyman}Here we list the 55~models that we have studied, each of them with its corresponding $\delta A$, namely the estimator of the small-scale power suppression associated to it. A model is excluded (at 95\%~C.L.) if $\delta A > \delta A_{\rm REF,1}$, i.e., if it shows a power suppression $\gtrsim 38\%$ with respect to the $\Lambda$CDM power spectrum. Each of the models corresponds to a different $\{\alpha, \beta, \gamma\}$-combination, according to Tab.~\ref{tab:params}. Accepted models are highlighted in bold-face.}
\end{center}
\end{table}
\begin{figure}[b]
  \centering
  \subfigure{\includegraphics[width=7.5cm]{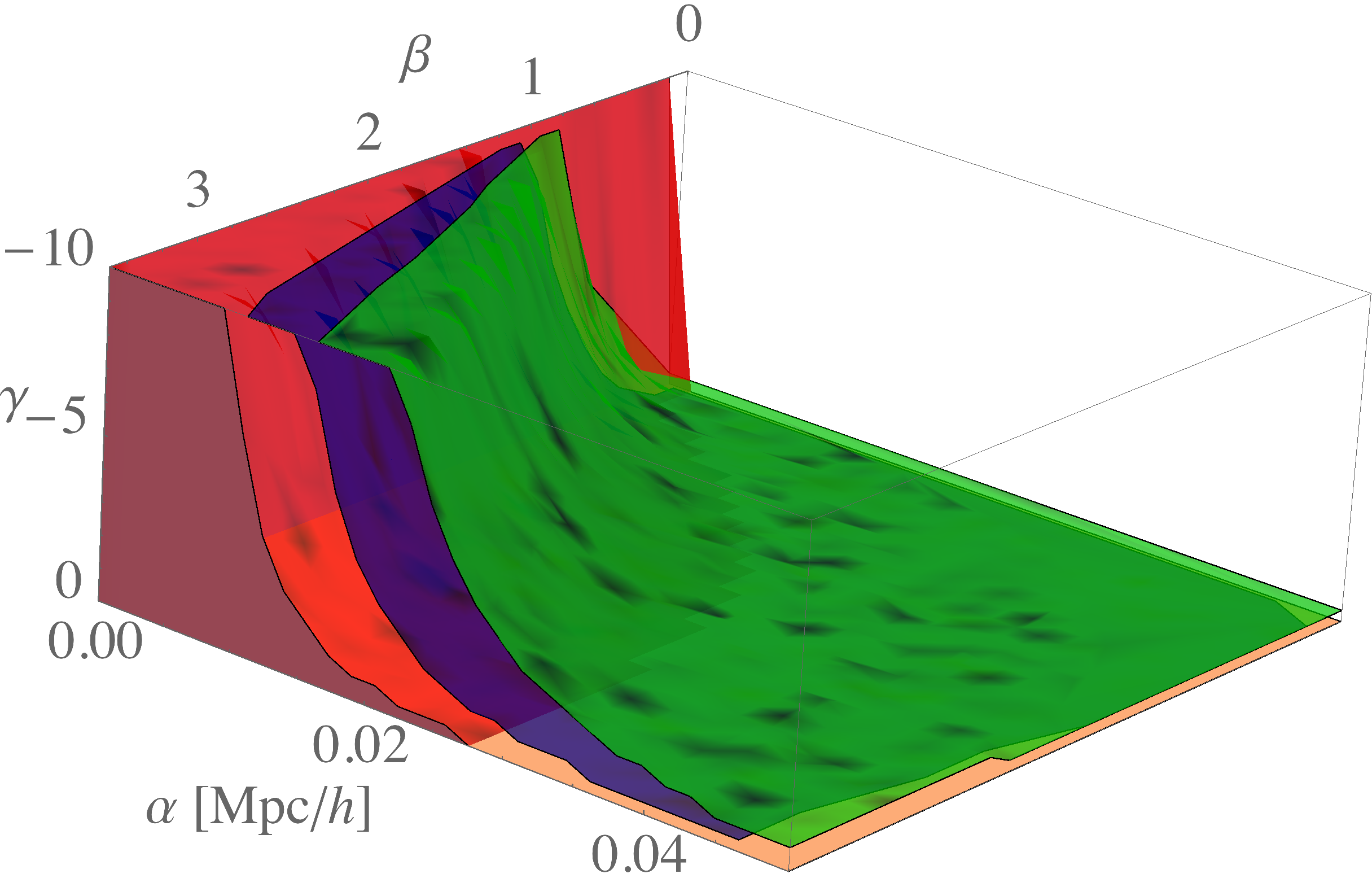}}
  \subfigure{\includegraphics[width=7.5cm]{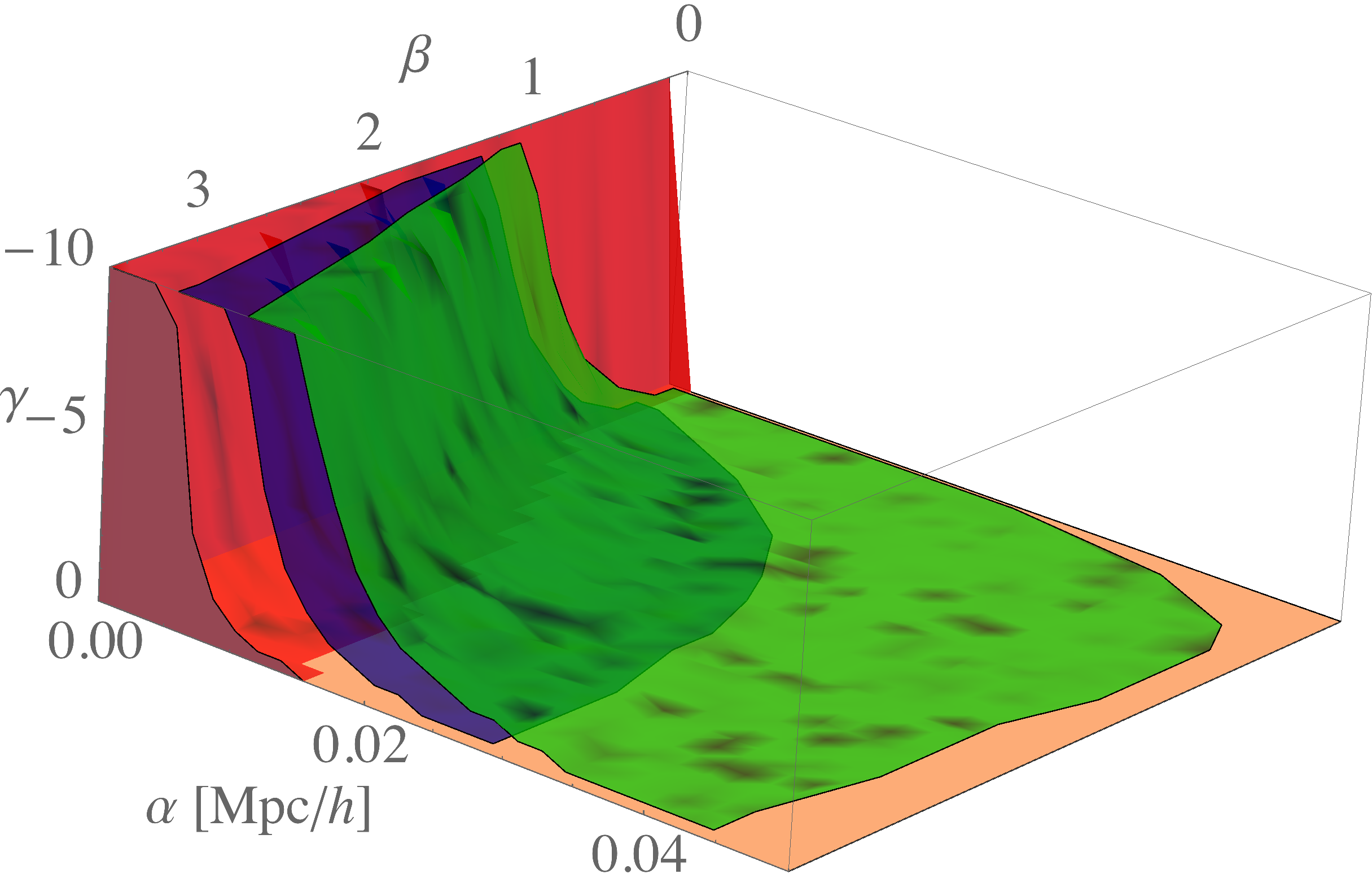}}
  \caption{\label{fig:lyman3d}\looseness=-1 The 3-dimensional contour plot in the $\{\alpha, \beta, \gamma\}$-space represents the region of the parameter space which contains models in agreement with the Lyman-$\alpha$ forest data. The left panel refers to the conservative analysis ($\delta A < 0.38$), whereas the right panel refer to the non-conservative case ($\delta A < 0.21$). The red contours represent the 68\%~C.L.\ limit on the $\{\alpha, \beta, \gamma\}$-combinations, while the blue and green contours represent the 95\% and 98\%~C.L.\ limits, respectively. All those models associated to $\{\alpha, \beta, \gamma\}$-triplets placed outside of the 3-dimensional coloured region are therefore excluded at 98\%~C.L.\ by our analyses. }
\end{figure} 

The Lyman-$\alpha$ forest (namely the Lyman-$\alpha$ absorption produced by intervening intergalactic neutral hydrogen in the spectra of distant quasars) provides a powerful tool for constraining small-scale DM properties. It is indeed well established that the Lyman-$\alpha$ absorption is produced by the inhomogeneous distribution of the Intergalactic Medium (IGM) along different line of sights to distant bright sources~\cite{Viel:2001hd}. Therefore, the Lyman-$\alpha$ forest has been used widely as a probe of the matter power spectrum on scales $0.5~{\rm Mpc}/h\, <  \lambda < 100~{\rm Mpc}/h\, $~\cite{Viel:2013apy,Baur:2015jsy,Irsic:2017ixq}.

\looseness=-1 Instead of computing absolute bounds, which can only be obtained through a full statistical analysis, our goal is to investigate the deviations of our nCDM models with respect to a thermal WDM reference model, i.e., $m_{\rm WDM} = 3.5$~keV, which is one of the most updated constraints on WDM candidates (at 95\%~C.L.), obtained through a recent comprehensive analysis of Lyman-$\alpha$ forest data~\cite{Irsic:2017ixq}. 

\looseness=-1 In order to constrain our models with Lyman-$\alpha$ forest data, we slightly modify the method developed in Ref.~\cite{Schneider:2016uqi}. We parametrise the deviation of a model with respect to $\Lambda$CDM by the ratio
\begin{equation}\label{eq:rk}
 r(k) = \frac{P_{1\rm{D}}(k)}{P^{\Lambda\rm{CDM}}_{1\rm{D}}(k)},
\end{equation}
where $P_{1\rm{D}}(k)$ is the 1D power spectrum of the model that we are considering, computed by the following integral on the 3D matter power spectrum:
\begin{equation}\label{eq:pk1d}
 P_{1\rm{D}}(k)=\frac{1}{2\pi} \int\limits_k^\infty {\rm d}k'k'P(k'),
\end{equation}
where $P(k')$ is the 3D linear matter power spectrum, computed at redshift $z=0$. 

\looseness=-1 We are now able to determine whether a model deviates more or less from $\Lambda$CDM, with respect to the thermal WDM reference model that we have chosen, by adopting the following criterion: a model is excluded (at 95\%~C.L.) if it is characterised  by a larger power suppression with respect to the reference model. In order to quantify the suppression in the power spectra, we define the following estimator:
\begin{equation}\label{eq:deltaA}
 \delta A \equiv \frac{A_{\rm \Lambda CDM} - A}{A_{\rm \Lambda CDM}},
\end{equation}
where $A$ is the integral of $r(k)$ over the typical range of scales probed by Lyman-$\alpha$ observations ($0.5~h/{\rm Mpc} < k < 20~h/{\rm Mpc}$ for the MIKE/HIRES+XQ-100 combined dataset used in~\cite{Irsic:2017ixq}), i.e.,
\begin{equation}\label{eq:A}
A = \int\limits_{k_{\rm min}}^{k_{\rm max}} {\rm d}k\ r(k),
\end{equation}
such that $A_{\rm \Lambda CDM} \equiv k_{\rm max} - k_{\rm min}$ by construction.

Analogously, by plugging the power spectrum of the thermal WDM reference model into Eqs.~\eqref{eq:rk} and~\eqref{eq:deltaA}, we find $\delta A_{\rm REF,1} = 0.38$, which is an estimate of the small-scale power suppression with respect to $\Lambda$CDM for models that are excluded at 95\%~C.L.\ by the Lyman-$\alpha$ forest data. In Tab.~\ref{tab:lyman} we list the 55~models that we have studied, each of them with its corresponding $\delta A$: a given model is excluded (at 95\%~C.L.) if $\delta A > \delta A_{\rm REF,1}$, i.e., if it shows a power suppression $\gtrsim 38\%$ with respect to the $\Lambda$CDM power spectrum. Accepted models are highlighted in bold-face.

Let us now stress that the constraint on the thermal WDM mass associated to $\delta A_{\rm REF,1}$ (i.e.\ $m_{\rm WDM} = 3.5$~keV) has been obtained under very conservative assumptions on the thermal history of the universe (see~\cite{Irsic:2017ixq} for details). By modifying these assumptions, the lower limit on thermal WDM masses strengthens to $m_{\rm WDM} = 5.3$~keV (at 95\%~C.L.), which represents the tightest bound from Lyman-$\alpha$ forest data up to date. By taking this limit as reference, we find indeed a corresponding small-scale suppression $\delta A_{\rm REF,2} = 0.21$, with respect to the $\Lambda$CDM power spectrum.

We note that the physical observable for the Lyman-$\alpha$ forest data is the \emph{flux power spectrum} $P_{\rm{F}}(k,z)$ and not the 1D or 3D linear matter power. However, two different key aspects of the Lyman-$\alpha$ forest physics suggest that the analysis presented could be also quantitatively correct. Firstly, the relation between linear matter and flux power can be modelled as $P_{\rm{F}}=b^2(k)P_{3\rm{D}}(k)$, with a bias factor $b^2(k)$ which differs only very little between $\Lambda$CDM and $\Lambda$nCDM models, at least for models reasonably close to the standard case (see e.g.~\cite{croft02,Viel2005}): this motivates the application of  Eq.~\eqref{eq:rk} to flux power spectra as well. Secondly, the area criterion is justified by the fact that IGM peculiar velocities (typically $<100$ km/s) tend to redistribute the small-scale power within a relatively wide range of wave-numbers in the probed region of the flux power~\cite{gnedin02}. Thereby, the derived bounds should be robustly checked against full hydrodynamic simulations which can provide a forward modeling of the flux power spectrum, as done in~\cite{Irsic:2017ixq}. We thus regard the approximate method presented here (based on the linear theory) as a first quantitative step towards a more comprehensive analysis.

In Fig.~\ref{fig:lyman3d} we show a 3-dimensional contour plot in the $\{\alpha, \beta, \gamma\}$-space, which represents the region of the parameter space that contains models in agreement with Lyman-$\alpha$ forest data. The left panel refers to the conservative analysis, whereas the right panel refers to the ``non-conservative'' case. The red contours represent the 68\%~C.L.\ limit on the $\{\alpha, \beta, \gamma\}$-combinations, while the blue and green contours represent the 95\% and 98\%~C.L.\ limits, respectively. All those models associated to $\{\alpha, \beta, \gamma\}$-triplets placed outside of the 3-dimensional coloured region are therefore excluded at 98\%~C.L.\ by our analysis. By marginalising over $\beta$ and $\gamma$ we obtain the following limits on $\alpha$:
\begin{equation}
 \begin{matrix}
 \alpha \leq 0.058~{\rm{Mpc}}/h~~(\rm{95\% ~C.L.}), & & & & & & & & & \text{conservative analysis,}\hfill\hfill\hfill\hfill\hfill\\
 \alpha \leq 0.044~{\rm{Mpc}}/h~~(\rm{95\% ~C.L.}), & & & & & & & & & \text{non-conservative analysis.}
 \end{matrix}
 \label{eq:alfalimit_lyman}
\end{equation}
\looseness=-1 These limits would correspond, in the old one-to-one parametrisation, to a thermal WDM particle with a mass of $m_{\rm{WDM}} \approx 3$~keV, see Eqs.~\eqref{eq:Viel} and~\eqref{eq:alphaold}. Even at this approximate level of analysis, the Lyman-$\alpha$ forest generally tend to provide more stringent constraints than MW satellite counts. Note that, however, this is not true for all models: e.g., the point nCDM35 is allowed by the Lyman-$\alpha$ forest, while it is excluded by satellite counts. This is because, after all, the two methods probe slightly different scales, and thus are in reality \emph{complementary} when it comes to constraining DM models. Note that this is a qualitative difference of non-thermal settings compared to thermal WDM: indeed, using different methods to constrain DM models is paramount to obtain a clear picture of what is allowed and what is not.
\looseness=-1 Still, the limit shown is weaker compared to the most updated constraints on thermal WDM masses. As we discussed before, this is due to the new general parametrisation of $T(k)$. With our approach, thanks to the mutual dependence among $\alpha$, $\beta$, and $\gamma$, it is possible to model nCDM scenarios with non-trivially suppressed power spectra. Therefore, models with shallower transfer functions may be found to be in agreement with Lyman-$\alpha$ forest data even if the corresponding DM candidate mass lies below the current constraints for thermal WDM masses, given that those constraints refer to a very specific shape of the power suppression (i.e., a very specific $\{\alpha, \beta, \gamma\}$-combination).

\looseness=-1 As a double-check, we applied our method to the non-linear power spectra extracted from the simulations, finding consistent results with respect to the linear analysis.
\looseness=-1 All the models rejected when comparing their linear power spectra are also rejected when comparing the non-linear ones.

\section{\label{sec:reality-check}Constraints on particle physics models}

The goal of this section is to compare predictions in terms of MW satellite counts and power in the range probed by the Lyman-$\alpha$ forest by using the fits and the true transfer functions of some particle physics models defined in Sec.~\ref{sec:models}, thus to put our fitting formula, Eq.~\eqref{eq:Tgen}, to the reality check.

\begin{table}[th]
\hspace{-0.5cm}
\begin{tabular}{|c||c|c|c|c||c|c||c|c|}
  \hline
& \small{$\alpha$} &  \small{$\beta$} & \small{$\gamma$} & \small{$k_{1/2}~[h/{\rm Mpc}]$} & \fn{$N^{\rm fit}_{\rm sub}$} (\fn{$N^{\rm true}_{\rm sub}$}) [{\purple \%}] & \fn{Agree?} & \fn{$\delta A_{\rm fit}$} (\fn{$\delta A_{\rm true}$}) [{\purple \%}] & \fn{Agree?} \\ \hline\hline
\scriptsize{} & \fn{$0.025$} & \fn{$2.3$} & \fn{$-2.6$} & \fn{17.276}& \fn{38 (39)} [{\purple $-2.6\%$}] & ${\CHECK}$ & \fn{0.555 (0.571)} [{\purple $-2.8\%$}] & ${\CHECK}$ \\
\scriptsize{\bf RP} & \fn{$0.071$} & \fn{$2.3$} & \fn{$-1.0$} & \fn{9.828} & \fn{15 (14)} [{\purple $+7.1\%$}] & ${\CHECK}$ & \fn{0.743 (0.754)} [{\purple $-1.5\%$}] & ${\CHECK}$ \\
\scriptsize{\bf neutrinos} & \fn{$0.038$} & \fn{$2.3$} & \fn{$-4.4$} & \fn{8.604} & \fn{5 (5)} [{\purple $\pm 0.0\%$}] & ${\CHECK}$ & \fn{0.799 (0.810)} [{\purple $-1.4\%$}] & ${\CHECK}$ \\
\scriptsize{} & \fn{$0.035$} & \fn{$2.1$} & \fn{$-1.5$} & \fn{15.073} & \fn{35 (37)} [{\purple $-5.4\%$}] & ${\CHECK}$ & \fn{0.599 (0.613)} [{\purple $-2.3\%$}] & ${\CHECK}$ \\
\hline                                                                                                                                                                                                                                                                                                                                                                                                                             
\scriptsize{\bf Neutrinos} & \fn{$0.016$} & \fn{$2.6$} & \fn{$-8.1$} & \fn{19.012} & \fn{38 (42)} [{\purple $-9.5\%$}] & ${\CHECK}$ & \fn{0.521 (0.535)} [{\purple $-2.6\%$}] & ${\CHECK}$ \\
\scriptsize{\bf from} & \fn{$0.011$} & \fn{$2.7$} & \fn{$-8.5$} & \fn{28.647} & \fn{{\bf 91} ({\bf 97})} [{\purple $-6.2\%$}] & ${\CHECK}$ & \fn{{\it 0.339} ({\it 0.360})} [{\purple $-5.8\%$}] & ${\CHECK}$ \\
\scriptsize{\bf particle} & \fn{$0.019$} & \fn{$2.5$} & \fn{$-6.9$} & \fn{16.478} & \fn{27 (28)} [{\purple $-3.6\%$}] & ${\CHECK}$ & \fn{0.582 (0.576)} [{\purple $+1.0\%$}] & ${\CHECK}$ \\
\scriptsize{\bf decay}	& \fn{$0.011$} & \fn{$2.7$} & \fn{$-9.8$} & \fn{26.31} & \fn{{\bf 79} ({\bf 87})} [{\purple $-9.2\%$}] & ${\CHECK}$ & \fn{0.375 ({\it 0.390})} [{\purple $-3.8\%$}] & {\FAIL} \\
\hline                                                                                                                                                                                                                                                                                                                                                                                                                                                        
\scriptsize{} & \fn{$0.16$} & \fn{$3.2$} & \fn{$-0.4$} & \fn{6.743} & \fn{9 (9)} [{\purple $\pm 0.0\%$}] & ${\CHECK}$ & \fn{0.823 (0.834)} [{\purple $-1.3\%$}] & ${\CHECK}$ \\
\scriptsize{\bf Mixed} & \fn{$0.20$} & \fn{$3.7$} & \fn{$-0.18$} & \fn{7.931} & \fn{28 (27)} [{\purple $+3.7\%$}] & ${\CHECK}$ & \fn{0.738 (0.752)} [{\purple $-1.9\%$}] & ${\CHECK}$ \\
\scriptsize{\bf models} & \fn{$0.21$} & \fn{$3.7$} & \fn{$-0.1$} & \fn{11.36} & \fn{{\it 60} ({\it 62})} [{\purple $-3.2\%$}] & ${\CHECK}$ & \fn{0.596 (0.610)} [{\purple $-2.3\%$}] & ${\CHECK}$ \\
\scriptsize{} & \fn{$0.21$} & \fn{$3.4$} & \fn{$-0.053$} & \fn{33.251} & \fn{{\bf 110} ({\bf 114})} [{\purple $-3.5\%$}] & ${\CHECK}$ & \fn{{\it 0.365} ({\it 0.377})} [{\purple $-3.2\%$}] & ${\CHECK}$ \\
\hline                                                                                                                                                                                                                                                                                                                                                                                                              
\scriptsize{} & \fn{$0.054$} & \fn{$5.4$} & \fn{$-2.3$} & \fn{13.116} & \fn{8 (9)} [{\purple $-11.1\%$}] & ${\CHECK}$ & \fn{0.691 (0.708)} [{\purple $-2.4\%$}] & ${\CHECK}$\\
\scriptsize{\bf Fuzzy} &  \fn{$0.040$} & \fn{$5.4$} & \fn{$-2.1$} & \fn{18.106} & \fn{21 (23)} [{\purple $-8.7\%$}] & ${\CHECK}$ & \fn{0.543 (0.565)} [{\purple $-3.9\%$}] & ${\CHECK}$\\
\scriptsize{\bf DM} & \fn{$0.030$} & \fn{$5.5$} & \fn{$-1.9$} & \fn{25.016} & \fn{56 (60)} [{\purple $-6.7\%$}] & ${\CHECK}$ & \fn{{\it 0.376} (0.399)} [{\purple $-5.8\%$}] & {\FAIL}\\
\scriptsize{} & \fn{$0.022$} & \fn{$5.6$} & \fn{$-1.7$} & \fn{34.590} & \fn{{\bf 121} ({\bf 126})} [{\purple $-4.0\%$}] & ${\CHECK}$ & \fn{{\it 0.228} ({\it 0.250})} [{\purple $-8.8\%$}] & ${\CHECK}$\\
\hline                                                                                                                                                                                                                                                                                                                                                                                                                                       
\scriptsize{} & \fn{$0.0072$} & \fn{$1.1$} & \fn{$-9.9$} & \fn{7.274} & \fn{18 (19)} [{\purple $-5.3\%$}] & ${\CHECK}$ & \fn{0.780 (0.788)} [{\purple $-1.0\%$}] & ${\CHECK}$\\
\scriptsize{\bf ETHOS} & \fn{$0.013$} & \fn{$2.1$} & \fn{$-9.3$} & \fn{16.880} & \fn{36 (39)} [{\purple $-7.7\%$}] & ${\CHECK}$ & \fn{0.568 (0.581)} [{\purple $-2.2\%$}] & ${\CHECK}$\\
\scriptsize{\bf models} & \fn{$0.014$} & \fn{$2.9$} & \fn{$-10.0$} & \fn{21.584} & \fn{50 (53)} [{\purple $-5.7\%$}] & ${\CHECK}$ & \fn{0.463 (0.477)} [{\purple $-2.9\%$}] & ${\CHECK}$\\
\scriptsize{} & \fn{$0.016$} & \fn{$3.4$} & \fn{$-9.3$} & \fn{23.045} & \fn{53 (56)} [{\purple $-5.4\%$}] & ${\CHECK}$ & \fn{0.430 (0.439)} [{\purple $-2.1\%$}] & ${\CHECK}$\\ \hline
\end{tabular}
\caption{\label{tab:final}\looseness=-1 Here we list 20 $\{\alpha, \beta, \gamma\}$-triplets, with the corresponding values of $k_{1/2}$, which represent the real model examples presented in Sec.~\ref{sec:models}, split into five groups. For each case, we have confronted the fit (the real model) with both halo counting, $N^{\rm fit}_{\rm sub}$ ($N^{\rm true}_{\rm sub}$), and the Lyman-$\alpha$ forest, $\delta A_{\rm fit}$ ($\delta A_{\rm true}$), where for each case we also indicate the percentage (in {\purple purple} for better visibility) by which the value predicted from the fit differs from the ``true'' value predicted by the model point. Bold-faced numbers indicate that both the restrictive and conservative bounds are met, while Italic numbers indicate that only the conservative bound was met. For each case we have indicated whether the conclusion drawn from the fit -- i.e., whether or not a certain choice of parameters is allowed by the data -- does (${\CHECK}$) or does not ({\FAIL}) agree with the one drawn from the data for the real model. Looking at the observables, it is visible that the values predicted by the fits and the real models differ by a few per cent at most and, indeed, the very few red crosses in the table indicate that our fitting function is remarkably powerful in reproducing the correct conclusion. Thus, in most cases, it is sufficient to fit a model with our general formula and check whether the fitted point is in the allowed region.}
\end{table}

\looseness=-1 Note that we do \emph{not} aim to give a full account of the validity (or invalidity) of the different nCDM models presented in Sec.~\ref{sec:models}, since in any case a few example points will not be able to give us a clear answer. Instead, we would like to find out whether the $\{\alpha, \beta, \gamma\}$-fit to a certain DM setting would lead us to the same conclusion about its validity when confronted with halo counting and Lyman-$\alpha$ bounds, while we do not care very much at this stage about whether a certain point in the model parameter space is now marked as ``excluded'' or ``allowed'' -- we only want to check whether our conclusion about the points under consideration changes when we look at the fits instead of looking at the actual points.

\looseness=-1 To do so, we depict in Tab.~\ref{tab:final} first of all the fit parameters for the example model points discussed in Sec.~\ref{sec:models}, which should be rather simple to grasp. For example, the second line for RP corresponds to the green curves in Fig.~\ref{fig:Res-fits}, while the first line for SD corresponds to the blue curves in Fig.~\ref{fig:SD-fits}. For each point, we have computed the number of satellites for both the fit and the corresponding transfer function of the real model, $N^{\rm fit}_{\rm sub}$ ($N^{\rm true}_{\rm sub}$), as well as the Lyman-$\alpha$ estimator, $\delta A_{\rm fit}$ ($\delta A_{\rm true}$) -- with the difference of the fit to the real point indicated by the percentages in square brackets -- which are in both cases matched to the respective conservative and non-conservative observational constraints. In all cases, no matter if the resulting number corresponds to a real model or to a fitted point, we use bold-faced scripts/Italic scripts/Roman scripts to indicate that a certain number is in agreement with both the conservative and non-conservative bounds/only with the conservative bound/with none of the bounds. As can be seen from the distribution of bold-faced or Italic numbers, many of the example points shown here are not in agreement with the bounds. However, what we are interested in is whether or not the fitted points would have brought us to the same conclusion. Indeed, this is the case for the vast majority of cases. In fact, given that the predictions from the fits deviate from the real model predictions only by a few per cent at most, we would expect agreement of the conclusions drawn from both versions of the transfer function (i.e., fitted and ``true'') in all cases up to a few per cent.\footnote{Two fails in fourty comparisons, i.e., an empirical failure rate of 5\% seems to support this prediction rather well -- even though, of course, we have not selected the examples shown completely arbitrarily, but rather we have picked them to somehow reflect some of the variation possible for the different production mechanisms.}

Thus, except for a tiny amount of borderline cases, the fitted points \emph{always} yield the same conclusion as the actual model points. This is the main result of our paper:
\begin{center}
\emph{Our fitting formula reproduces the true results to a very high degree.}
\end{center}
Hence, whenever is it desired to match a nCDM setting to observational data, there is no need to do the whole computation. Instead, it is perfectly sufficient to match the resulting transfer functions to our Eq.~\eqref{eq:Tgen} and to check whether the fitted points are allowed -- which is known from the present text already, and which can for sure be improved in future studies. In fact, one could in principle also constrain the parameter space of our fitting formula by elaborate full scale $N$-body simulations, which would to quite some extent get rid of the potential necessity to run a computationally expensive simulation for a quasi-infinite list of models. With such constraints available, our fit could be used to easily identify the interesting regions in the parameter spaces of many types of complicated particle physics models of non-thermal DM, by simply computing linear power spectra.

\section{\label{sec:concl}Conclusions}

Various dark matter scenarios predict structure formation to be suppressed at small cosmological scales. This suppression can be of different strength and shape depending on the particle nature of the DM candidate. In the past, a lot of effort has gone into investigating the astrophysical consequences of thermal WDM (i.e.~candidates with a Fermi-Dirac or Bose-Einstein momentum distribution), which is characterised by a strong suppression of power only depending on the WDM particle mass. While being a good example for a ``non-cold'' scenario, the case of thermal WDM is far too restrictive to give justice to the large variety of potential nCDM models.

\looseness=-1 In this paper we develop a new method to accurately capture the clustering of nCDM models, based on a generalised fit of the transfer function (i.e., the square root of the linear power spectrum relative to the case of pure cold DM). The fit, given by Eq.~\eqref{eq:Tgen}, is based on a set of 3 free parameters, regulating the scale and the shape of the power suppression in a very flexible way. 
\looseness=-1 Our approach is meant to provide a connection between DM model building and astrophysics. It is expected to work efficiently for most of the known nCDM models. We have explicitly shown that this approach covers the full parameter space of sterile neutrinos (both by resonant production and particle decays), mixed (cold and warm) models, fuzzy DM, and models suggested by ETHOS. Due to the mutual dependence among the three free parameters of the new fitting formula, we are now able to disentangle even tiny differences in the shape of the power suppression and thus to distinguish between models with power spectra suppressed at very similar scales, in order to test them individually with cosmic structure formation data.

We have performed a large suite of $N$-body DM-only simulations that conservatively bracket the suppression of power suggested by the constraints obtained up to date from the Lyman-$\alpha$ forest and we have extracted the non-linear matter power spectra and (sub)halo mass functions.

\looseness=-1 We have given the first astrophysical constraints on the three free parameters which characterise our method, by using two independent astrophysical observables: the number of MW satellites and the Lyman-$\alpha$ forest. From this preliminary analysis it appears that no significant suppression of power can be present at scales larger than $\sim0.06~h^{-1}$ (comoving Mpc) in order to fit both the observables, regardless of the shape of the power spectrum at smaller scales. This result is valid under the framework of abrupt suppressions of power happening at or below the scales $k\sim 1\, h$/Mpc, those typically induced by nCDM models, rendering the more gentle neutrino-induced suppression at larger scales still possible.

Finally we have constrained some of the most popular nCDM scenarios, by fitting the corresponding transfer functions with the new general formula and then testing them with MW satellite counts and Lyman-$\alpha$ forest data, using the methods that we have outlined.

This work represents a first step towards a general and comprehensive modeling of small-scale departures from the standard CDM paradigm. Future developments will include a full statistical analysis of Lyman-$\alpha$ forest data, by performing hydrodynamical simulations in order to extract the flux power spectra for our nCDM scenarios and determine more accurate limits on $\{\alpha, \beta, \gamma\}$, and a weak lensing data analysis, which will provide another independent observable for constraining the parameter space.
It is also expected that forthcoming instruments like DESI\footnote{http://desi.lbl.gov/} or the Euclid satellite\footnote{https://www.euclid-ec.org/} 
could help in constraining the parameters by exploiting the weak lensing signal and or the clustering of galaxies at small scales, even if it is likely that most of the progress will be made by combining the DESI quasar spectroscopic survey with higher resolution Lyman-$\alpha$ forest spectra. Another interesting possibility, featuring a new observational probe, will be to probe the global intensity mapping signal of the 21 cm transition produced by neutral hydrogen, which has the advantages of being at high redshift where non-linear evolution is less important~\cite{carucci15}.

\section*{Acknowledgments}

We would like to thank Francis-Yan Cyr-Racine for providing the data used in Fig.~1a of Refs.~\cite{Cyr-Racine:2015ihg,Vogelsberger:2015gpr}, and we are also grateful for Nicola Menci for useful discussions on the small-scale part of the transfer functions. AM acknowledges partial support by the Micron Technology Foundation, Inc. AM furthermore acknowledges partial support by the European Union through the FP7 Marie Curie Actions ITN INVISIBLES (PITN-GA-2011-289442) and by the Horizon 2020 research and innovation programme under the Marie Sklodowska-Curie grant agreements No.~690575 (InvisiblesPlus RISE) and No.674896 (Elusives ITN). RM and MV are supported by the INFN PD51 grant INDARK. MV is also supported by the ERC-StG ``cosmoIGM''. 
AS acknowledges support from the Ambizione grant of the Swiss National Science Foundation (project number PZ00P2$\_$161363).
Simulations were performed on the Ulysses SISSA/ICTP supercomputer. MT acknowledges support by Studienstiftung des deutschen Volkes as well as support by the IMPRS-EPP.

\appendix
\section{\label{ap:degeneracy}Quasi-degeneracy between $\boldsymbol{\alpha}$ and $\boldsymbol{\gamma}$}

By considering the high-$k$ limit of Eq.~\eqref{eq:Tgen}, we can easily notice the quasi-degeneracy between the two parameters $\alpha$ and $\gamma$. For large $k$ (i.e., small scales), such that $(\alpha k)^\beta \gg 1$, we have indeed:
\begin{equation}
 \left.T \left(k\right)\right|_{k \alpha \gg 1} \simeq \alpha^{\beta \gamma} k^{\beta \gamma}.
 \label{eq:deg_1}
\end{equation}
If we now change $\alpha$ to a new value $\alpha \to \tilde \alpha = x \cdot \alpha$, where $x$ is some real number, we can absorb this changes into a $k$-dependent change of $\gamma$:
\begin{equation}
 \alpha^{\beta \gamma} k^{\beta \gamma} \to \tilde \alpha^{\beta \gamma} k^{\beta \gamma} = \alpha^{\beta \gamma} x^{\beta \gamma} k^{\beta \gamma} = \alpha^{\beta \gamma} \left(k^{\ln x/\ln k}\right)^{\beta \gamma} k^{\beta \gamma} = \alpha^{\beta \gamma} k^{\beta \gamma (1 + \ln x/\ln k)},
 \label{eq:deg_2}
\end{equation}
where we have used the obvious identity $x = k^{\ln x/\ln k}$. Thus, in Eq.~\eqref{eq:deg_1}, we can trade a change in $\alpha$ for a $k$-dependent change in $\gamma$:
\begin{equation}
 \gamma \to \tilde \gamma (k) = \gamma (1 + \ln x/\ln k),
 \label{eq:gamma-change}
\end{equation}
which reproduces the effect of the change from $\alpha$ to $\tilde \alpha$. Now, this does not seem like a real degeneracy -- and mathematically it is not, due to the $k$-dependence of $\tilde \gamma$. However, this dependence is only very weak (logarithmic), so that one can, in spite of the $k$-dependence, view $\tilde \gamma$ as approximately constant:
\begin{equation}
 \tilde \gamma (k) \simeq {\rm const.}
 \label{eq:gamma-constant}
\end{equation}
This approximation will be even better if $|\log_k x| = |\ln x/\ln k| \ll 1$, due to this quantity only appearing in the sum in Eq.~\eqref{eq:gamma-change}. Thus, indeed, to a very good approximation there is a degeneracy between $\alpha$ and $\gamma$: a change in one of them can be traded for a change in the other. Hence, by covering several different values of $\gamma$ in Fig.~\ref{fig:grid}, we have information about many different $\alpha$'s at the same time.

\section{\label{ap:numconv}Comparing simulated halo mass functions with theoretical predictions}
\begin{figure}[t]
\begin{tabular}{lr}
  \includegraphics[page=1,width=7.5cm]{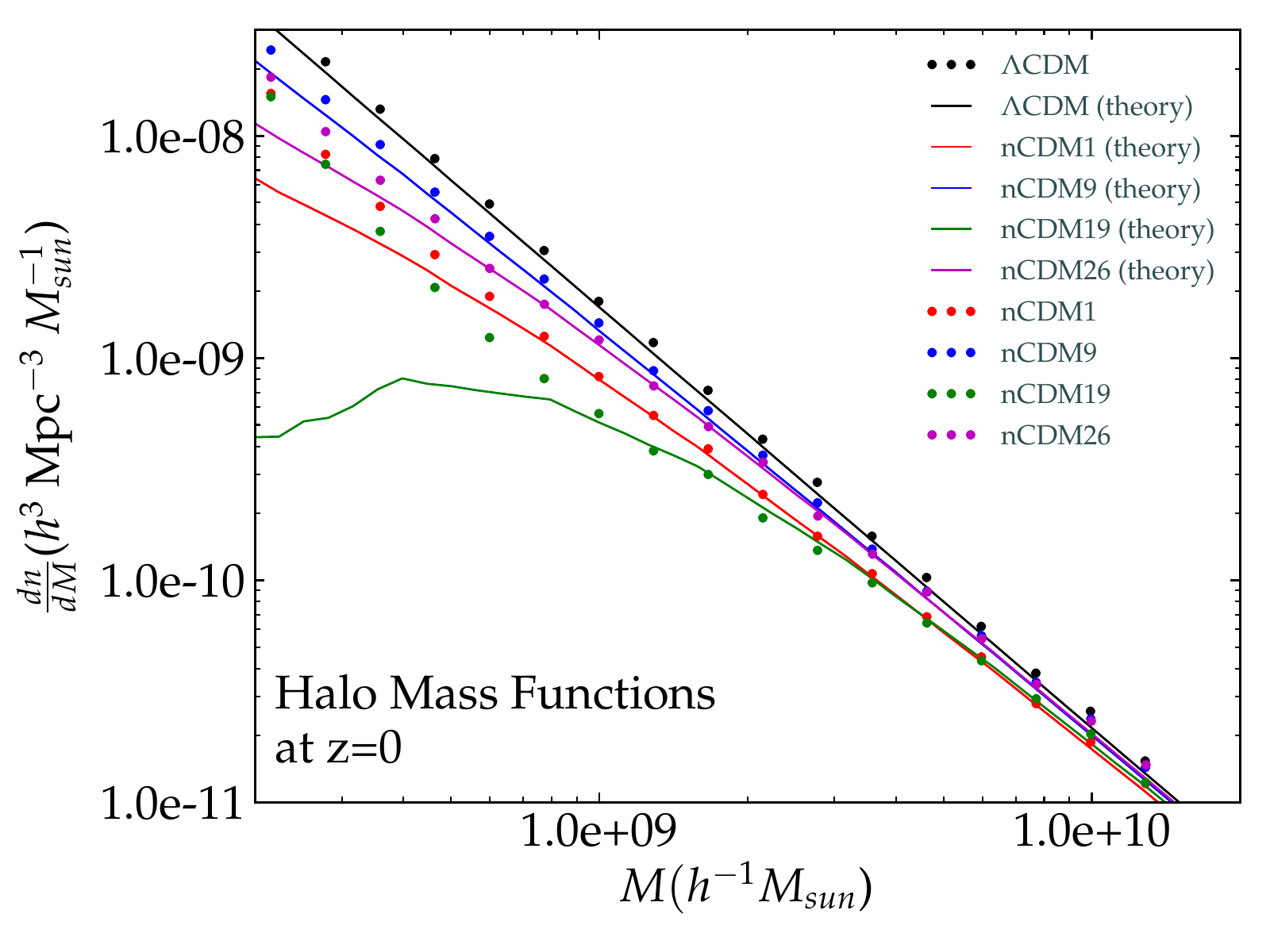} & \includegraphics[page=1,width=7.5cm]{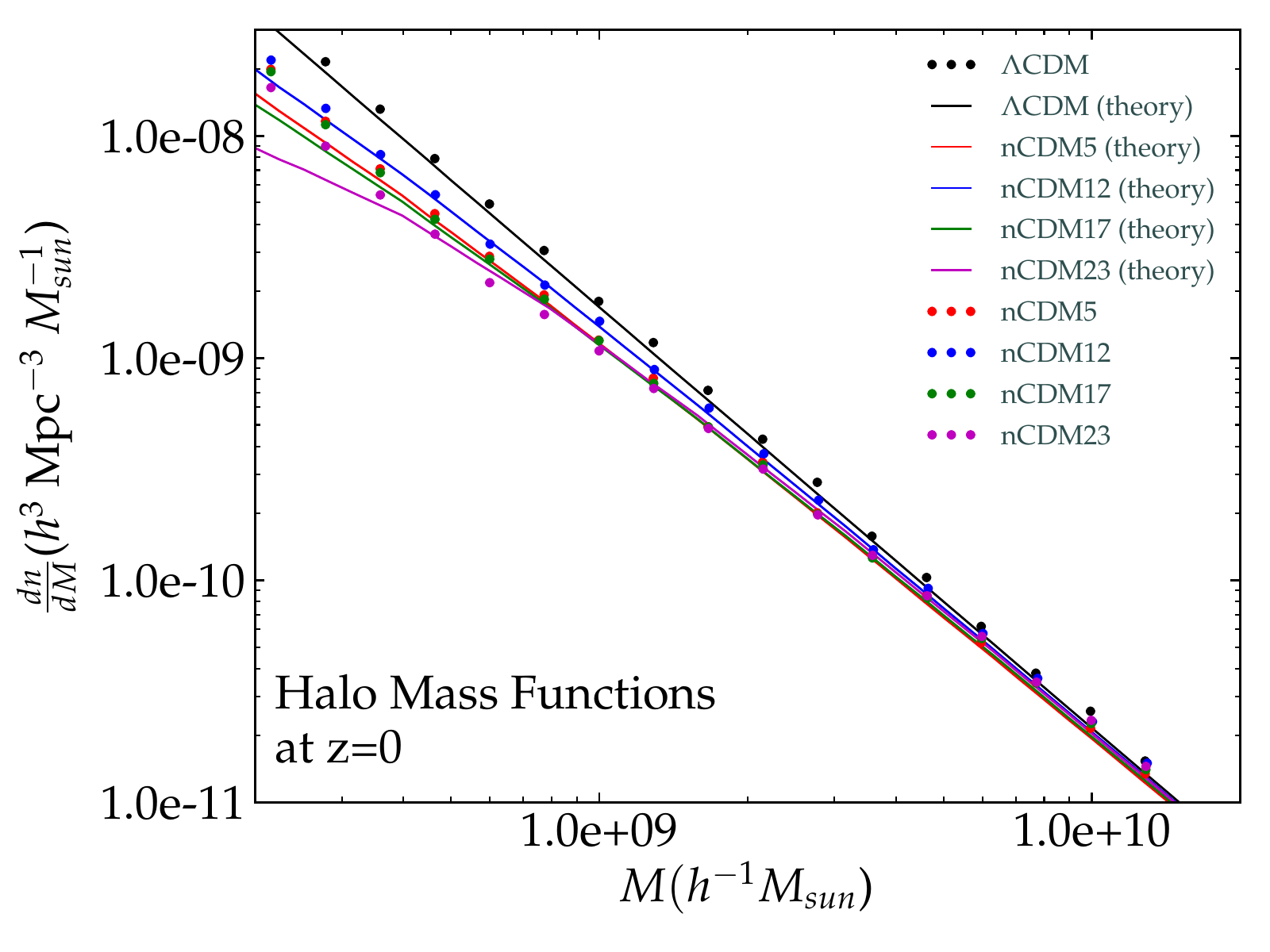}
  \end{tabular}
\caption{\label{fig:MF_simsVSth}For some of the models that we have studied, we compare the theoretical halo mass function predicted by Eq.~\eqref{eq:MF} with the halo mass function extracted from the corresponding $N$-body simulation. The former is plotted as a solid line, the latter as a dotted one. Different colours refer to different models. The good agreement between them ceases to hold below masses of the order of $10^9~M_\odot/h$, where artificial clumping strongly affects the results of the simulations. That mass indeed corresponds to the upturn highlighted in Fig.~\ref{fig:hmf}.}
\end{figure}

We have already highlighted in Sec.~\ref{subsec:hmf} the low-mass upturn in the halo mass functions, due to artificial clumping. To take this effect into account and subtract the corresponding numerical artefacts, we have estimated the number of subhalos predicted by our models by integrating over Eq.~\eqref{eq:subMF}, which is the theoretical subhalo mass function derived by an extended Press-Schechter approach~\cite{1974ApJ187425P,Sheth:1999mn} based on a sharp-$k$ window~\cite{Bertschinger:2006nq,Benson:2012su}. This leads to the following expression~\cite{Schneider:2013ria,Schneider:2014rda}:
\begin{equation}\label{eq:MF}
 \frac{{\rm d}n}{{\rm d}M} = \frac{1}{12\pi^2}\frac{\overline{\rho}}{M^2}\nu f(\nu) \frac{P(1/R)}{\delta^2_cR^3},
\end{equation}
\looseness=-1 where $\overline{\rho}$ is the average density of the universe, $\nu = \delta_{c,0}^2/S(R)$ is the peak-height of the perturbations (at $z=0$), and $f(\nu)$ is obtained by the excursion-set approach~\cite{Bond:1990iw}. The relation between the sharp-$k$ filter scale and mass and the variance $S(R)$ are defined in Eq.~\eqref{eq:mass_var}.
\looseness=-1 In Fig.~\ref{fig:MF_simsVSth}, we compare example theoretical halo mass function predicted in Eq.~\eqref{eq:MF} to the halo mass function extracted from the corresponding $N$-body simulations. The former is plotted as a solid line, the latter as a dotted line. Consistenly with the trend outlined in Fig.~\ref{fig:hmf}, below masses of the order of $10^9~M_\odot/h$, artificial clumping induces the upturn in the simulated halo mass functions and hence the discrepancies with the theory which are manifest in Fig.~\ref{fig:MF_simsVSth}, below masses of the order of $10^9~M_\odot/h$.

\bibliographystyle{utcaps}
\bibliography{wdm}

\end{document}